\documentclass[10pt,a4paper]{article}
\usepackage[utf8]{inputenc}
\usepackage[english]{babel}

\usepackage{amsmath}
\usepackage{amsfonts}
\usepackage{amssymb}

\usepackage{amsthm}

\usepackage[colorlinks,citecolor=blue,urlcolor=blue,linkcolor=blue]{hyperref}

\usepackage[left=2cm,right=2cm,top=2cm,bottom=2cm]{geometry}

\usepackage{authblk}

\usepackage{feynmp}
\newcommand{\pd}{\partial}
\newcommand{\okappa}[1]{\mathcal{O}\left( \kappa^{#1} \right)}

\title{FeynGrav 3.0}
\author[1]{\href{https://orcid.org/0000-0001-7099-0861}{Boris Latosh} \thanks{ \href{mailto:latosh.boris@ibs.re.kr}{latosh.boris@ibs.re.kr} } }
\affil[1]{Particle Theory  and Cosmology Group, Center for Theoretical Physics of the Universe, Institute for Basic Science (IBS), Daejeon, 34126, Korea}
\date{CTPU-PTC-24-15}

\begin{document}

\maketitle

\begin{abstract}
  We present the new version of the \texttt{FeynGrav}. The package provides tools to operate with Feynman rules for quantum gravity within \texttt{FeynCalc}. The latest version improves package efficiency and implements new physical models. We discover recurrent relations between metric factors that enhance computational efficiency. We discuss gravitational interaction with Horndeski gravity, quadratic gravity, and the simplest axion-like coupling. We implemented the massive graviton propagator and discussed the possibility of implementing massive gravity within the package.
\end{abstract}

\section{Introduction}

This paper discussed the new version of \texttt{FeynGrav} \cite{FeynGrav,Latosh:2022ydd,Latosh:2023zsi,Latosh:2024anf}, a \texttt{Wolfram Mathematica} package based on \texttt{FeynCalc} \cite{Mertig:1990an,Shtabovenko:2016sxi,Shtabovenko:2020gxv,Shtabovenko:2021hjx}, providing a set of tools to operate with Feynman rules within perturbative quantum gravity.

Perturbative quantum gravity is the simplest approach to quantum gravity based on the standard quantum field theory. The view on the perturbative quantum gravity is twofold. Firstly, one can treat it as a quantum field theory of small metric perturbations $h_{\mu\nu}$ that exist on a flat background \cite{Latosh:2022ydd,Latosh:2023zsi,Latosh:2024anf,Iwasaki:1971vb,Donoghue:1994dn,Burgess:2003jk,Accioly:2000nm,Calmet:2013hfa,Prinz:2018dhq,Prinz:2020nru}. In that case, one composes the complete spacetime metric $g_{\mu\nu}$ from the background metric $\eta_{\mu\nu}$ and the perturbations $h_{\mu\nu}$:
\begin{align}
  g_{\mu\nu} \overset{\text{def}}{=} \eta_{\mu\nu} + \kappa \, h_{\mu\nu}.
\end{align}
Here $\kappa$ is a gravitational coupling with the mass dimension $-1$, which ensures that the perturbations have canonical mass dimension. The coupling is related to the Newton's constant $G_\text{N}$:
\begin{align}
  \kappa^2 = 32\,\pi\,G_\text{N}.
\end{align}
Although the metric $g_{\mu\nu}$ is a finite expression, it gives rise to infinite series. For instance, the inverse metric $g^{\mu\nu}$ is an infinite series in $\kappa$. The volume factor $\sqrt{-g}$ is also an infinite series due to the presence of the square root. Similarly, the Riemann tensor $R_{\mu\nu\alpha\beta}$ and the Christoffel symbols $\Gamma^\alpha_{\mu\nu}$ involve the inverse metric, so they are infinite series as well. The resulting theory admits infinite interaction terms, all suppressed by different powers of the same gravitational coupling. 

Secondly, one can view perturbative quantum gravity as a gauge theory of massless particles with chirality $\pm 2$ described by the standard quantum field theory. In turn, quantum field theory admits many powerful tools that constrain the structure of the theory. For instance, gauge invariance constrains the form of interaction rules, while the optical theorem recovers a point of scattering amplitudes. These tools allow one to retrieve information about graviton propagator, interaction vertices, and certain scattering amplitudes \cite{Fierz:1939ix,rivers1964lagrangian,Weinberg:1964ew,Weinberg:1965nx,VanNieuwenhuizen:1973fi,Elvang:2015rqa,Arkani-Hamed:2017jhn,Travaglini:2022uwo}. These approaches are equivalent for two reasons. First, they recover the same results with different techniques. Secondly, they use other means of description to operate with the same physical setup. The flat background in the first approach ensures that one can consistently define the asymptotic states respecting the Poincare symmetry. The presence of the Poincare symmetry, in turn, allows one to define states with definite mass and spin/chirality \cite{Wigner:1939cj}. This setup allows one to use the formalism of quantum field theory. In turn, the theory invariance with respect to the change of coordinates ensures an infinite number of interaction terms. Within the first approach, the coordinate invariance of action requires the presence of the volume factor $\sqrt{-g}$ to preserve the coordinate volume invariant of the inverse metric $g^{\mu\nu}$ to contract covariant derivatives and other geometric quantities which are infinite perturbative series. Within the second approach, the coordinate invariance is a gauge symmetry. The gauge invariance of scattering matrix elements ensures infinite graviton interaction terms exist.

It is essential to discuss the physical features of the perturbative quantum gravity, but it is not the main scope of this article. We highlighted the main features of the theory above and omitted further discussion. Papers \cite{Latosh:2024anf,Donoghue:1994dn,Burgess:2003jk,Calmet:2013hfa} discuss the theory's applicability domain and predictive ability in more detail.

The new version of \texttt{FeynGrav} implements the following features. First and foremost, it implements the recently discovered recursive relations. It is possible to obtain explicit formulas for many geometric quantities \cite{Latosh:2022ydd}. However, they typically involve multiple summations, which affects the performance. The recursive relations simplify such summations and improve the computational performance. 

Secondly, the new version implements Feynman rules for a new series of models. First and foremost, we address the Horndeski gravity. One can obtain the interaction rules for all Horndeski models within the perturbative quantum gravity. It was firstly discussed in \cite{Latosh:2024anf}, and we extend this discussion below. On the practical ground, most parts of Horndeski's models that involve the coupling of many scalars are irrelevant. Consequently, the new \texttt{FeynGrav} version implements the interaction rules only for a few essential cases. 

Further, the new version implements the massive graviton propagator. Massive gravity on its own is a separate branch of research reached with phenomenology \cite{deRham:2014zqa,Hinterbichler:2011tt}. It is desirable to obtain a similar algorithm deriving the interaction rules for any order in perturbation theory, but the task faces a significant challenge. Due to the non-vanishing graviton mass, the theory does not admit the gauge symmetry, which results in particular challenges. We discuss this issue in detail below and will address it in further publications. 

The new version implements the simplest case of the axion-like coupling between a scalar field and a single vector field. The term describes a contraction of a vector field tensor $F$, and the dual tensor $\tilde{F}$ takes a special place in particle physics. The term is a complete derivative, so it does not contribute to the classical field equations. Within the $SU(N)$ Yang-Mills theory, it is known as the $\theta$ term. It is responsible for the existence of instantons and the related phenomenology \cite{tHooft:1976snw,Wilczek:1977pj,Weinberg:1977ma,Dine:1981rt,Dine:1982ah,Preskill:1982cy,Sikivie:1983ip}. The need to explain the value of the $\theta$ term produced various axion models that involve a coupling of a scalar field to the $F\tilde{F}$ term. This publication does not address the coupling for the $SU(N)$ Yang-Mills case since it deserves a separate publication. This paper considers the simplest case when a scalar field is coupled to the $F\tilde{F}$ term. Despite its simplicity, interaction is essential for many models and provides an intermediate step in developing \texttt{FeynGrav}.

Lastly, we address the quadratic gravity studied in classical papers \cite{Stelle:1976gc,Stelle:1977ry,Julve:1978xn,Fradkin:1981iu,Elizalde:1994sn,Elizalde:1994av,Lavrov:2019nuz}. We do not discuss the physical content of the model and do not address the issues related to its applicability. We focus on the derivation of the Feynman rules of the model. Their derivation follows the same principal scheme and only uses more sophisticated expressions for higher curvature invariants.

The paper is organised as follows. In Section \ref{Section_Recursive}, we discuss the recursive relations between various geometrical quantities. Section \ref{Section_Horndeski} discusses the Horndeski gravity, and it is split into subsections for each class of Horndeski interactions. We discuss the coupling to the Gauss-Bonnet term in Section \ref{Section_Gauss_Bonnet}. Within the standard Horndeski parametrisation, it involves the $\ln$ function of a scalar field, which makes it inconvenient to use. We address this issue by studying this term in a more convenient parametrisation. Section \ref{Section_massive_gravity} addresses the massive graviton propagator and discusses the applicability of \texttt{FeynGrav} for massive gravity. Section \ref{Section_axion} discusses the axion-like coupling between a scalar field and a single vector field. Section \ref{Section_Quadratic_Gravity} discusses the interaction rules in quadratic gravity. Lastly, Section \ref{Section_V3} discusses all the new features implemented in the latest version of \texttt{FeynGrav}. Section \ref{Conclusions} presents our conclusions and discusses further development of \texttt{FeynGrav}.

\section{Recursive relations}\label{Section_Recursive}

This section discusses recursive relations that hold for the perturbative expansion about the flat spacetime. We are introducing a new notation system that is better suited to our purpose. New notations are required to differentiate between the tensors that define the structure of perturbative expansions and those that represent the Feynman rules. The tensors in perturbative expansions do not possess symmetry unless explicitly stated, so we refer to them as plain tensors. We do not reserve a particular name for the tensors present in the interaction rules but use cursive symbols to denote them.

The metric perturbative expansion is defined as follows:
\begin{align}
  g_{\mu\nu} \overset{\text{def}}{=} \eta_{\mu\nu} + \kappa\, h_{\mu\nu}.
\end{align}
Here $\kappa$ is the gravitational coupling with mass dimension $-1$ related to the Newton constant $G_\text{N}$.
\begin{align}
  \kappa^2 \overset{\text{def}}{=} 32 \, \pi \, G_\text{N}.
\end{align}

The plain $I_{(n)}$-tensor of the $n$-th order has $2 n$ indices and defined as follows:
\begin{align}
  I^{\rho_1\sigma_1\cdots\rho_n\sigma_n}_{(n)} \overset{\text{def}}{=} \eta^{\sigma_1\rho_2} \eta^{\sigma_2\rho_3}\cdots \eta^{\sigma_n\rho_1}.
\end{align}
As stated above, this tensor does not have additional symmetries.

The $\mathcal{I}_{(n)}$-tensor of the $n$-th order has $2 n$ indices are defined as follows:
\begin{align}
  \mathcal{I}_{(n)}^{\rho_1\sigma_1\cdots\rho_n\sigma_n} \overset{\text{def}}{=} \cfrac{1}{2^n} \, \cfrac{1}{n!} \left[ I_{(n)}^{\rho_1\sigma_1\cdots\rho_n\sigma_n} + \text{permutations} \right].
\end{align}
The permutation term ensures that the $\mathcal{I}$-tensor has additional symmetries. It is constructed to be symmetric with respect to permutations of indices within each index pair $\rho_i\leftrightarrow\sigma_i$, and with respect to permutations of any two index pairs $\{\rho_i,\sigma_i\} \leftrightarrow \{\rho_j,\sigma_j\}$.

The number of terms present in a tensor of the $n$-th order is $n! \,2^n$. These terms are due to the tensor symmetry, so they cannot be negated. Consequently, the expression for interaction vertices will also have a rapidly growing number of terms, so the complexity of calculations will immensely grow with each order of perturbation theory. This feature is expected since the perturbative quantum gravity is an effective theory.

The plain $I$-tensors completely define the $g^{\mu\nu}$ perturbative expansion:
\begin{align}
  g^{\mu\nu} &= \sum\limits_{n=0}^\infty (-1)^n\,\kappa^n\,I^{\mu\nu\rho_1\sigma_1\cdots\rho_n\sigma_n}_{(1+n)}\,h_{\rho_1\sigma_1}\cdots h_{\rho_n\sigma_n}\,.
\end{align}
The reason is the structure of the $I$-tensor. By the construction, the tensor is designed to contract indices and form traces:
\begin{align}
  \begin{split}
    &(h^n)^{\mu\nu}  = I_{(1+n)}^{\mu\nu\rho_1\sigma_1\cdots\rho_n\sigma_n}\,h_{\rho_1\sigma_1}\cdots h_{\rho_n\sigma_n}\,,\\
    & \operatorname{tr}(h^n) =I_{(n)}^{\rho_1\sigma_1\cdots\rho_n\sigma_n}\,h_{\rho_1\sigma_1}\cdots h_{\rho_n\sigma_n}\,.
  \end{split}
\end{align}
Therefore, the plain $I$ tensor is sufficient to completely encapsulate the inverse metric's perturbative structure.

The plain $C$-tensor of $n$-th order has $2n$ indices and describes the structure of the volume factor $\sqrt{-g}$:
\begin{align}
  \sqrt{-g} \overset{\text{def}}{=} \sum\limits_{n=0}^\infty \, \kappa^n \,C^{\rho_1\sigma_1\cdots\rho_n\sigma_n}_{(n)}\,h_{\rho_1\sigma_1}\cdots h_{\rho_n\sigma_n} \,.
\end{align}
This definition is non-constructive since we must describe how to generate the tensor. The publication \cite{Latosh:2022ydd} presents an explicit formula for the plain $C$-tensor, but it requires a complicated summation. We will proceed to obtain a recurrent relation defining this tensor.

In full analogy with the previous case, the plain $C$-tensor does not have additional symmetries. One shall also define its symmetric counterpart. The $\mathcal{C}$-tensor of the $n$-th order in the following way:
\begin{align}
  \mathcal{C}_{(n)}^{\rho_1\sigma_1\cdots\rho_n\sigma_n} \overset{\text{def}}{=} \cfrac{1}{2^n} \, \cfrac{1}{n!} \, \left[ C_{(n)}^{\rho_1\sigma_1\cdots\rho_n\sigma_n} + \text{permutations} \right].
\end{align}
Permutations account for all terms that make the $\mathcal{C}$-tensor symmetric with respect to permutations of indices within each index pair $\rho_i\leftrightarrow\sigma_i$ and with respect to permutations of any two index pairs $\{\rho_i,\sigma_i\} \leftrightarrow \{\rho_j,\sigma_j\}$.

The following relation allows one to define the plain $C$-tensor completely
\begin{align}\label{the_C_recursion}
  C_{(n)}^{\rho_1\sigma_1\cdots\rho_n\sigma_n} = \cfrac{1}{2\,n} \sum\limits_{k=1}^n \,(-1)^{k-1} \, I_{(k)}^{\rho_1\sigma_1\cdots\rho_k\sigma_k} C_{(n-k)}^{\rho_{k+1}\sigma_{k+1}\cdots\rho_n\sigma_n}.
\end{align}

This relation is a consequence of a scaling behaviour. Let us define a new metric with scaled perturbations:
\begin{align}
  \mathbf{g}_{\mu\nu} = \eta_{\mu\nu} + z\, \kappa\,h_{\mu\nu}\,.
\end{align}
In this expression, $z$ is an arbitrary positive real number that scales the perturbation amplitude. This metric has no additional physical or mathematical meaning; we only use it to derive the relations discussed. 

The infinitesimal action of this scaling on the volume factor is
\begin{align}
  \left. \cfrac{d}{dz} \sqrt{-\mathbf{g}} ~ \right|_{z=1} ,
\end{align}
and we have two ways to calculate this quantity. Firstly, we can use the Jacobi's formula:
\begin{align}
  \left. \cfrac{d}{dz} \sqrt{-\mathbf{g}} ~ \right|_{z=1} = \cfrac{\kappa}{2} \, \sqrt{-g} \,g^{\mu\nu} \, h_{\mu\nu}.
\end{align}
Secondly, we can use the formula for the explicit perturbative structure:
\begin{align}
  \left. \cfrac{d}{dz} \sqrt{-\mathbf{g}} ~ \right|_{z=1} = \sum\limits_{n=1}^\infty \,n\,\kappa^n\,C^{\rho_1\sigma_1\cdots\rho_n\sigma_n}_{(n)}\,h_{\rho_1\sigma_1}\cdots h_{\rho_n\sigma_n}.
\end{align}
These equations allows us to exclude $\mathbf{g}$ completely:
\begin{align}
  \cfrac{\kappa}{2} \, \sqrt{-g} \,g^{\mu\nu} \, h_{\mu\nu} = \sum\limits_{n=1}^\infty \,n\,\kappa^n\,C^{\rho_1\sigma_1\cdots\rho_n\sigma_n}_{(n)}\,h_{\rho_1\sigma_1}\cdots h_{\rho_n\sigma_n}.
\end{align}
The left-hand side of this expression can also be calculated as a perturbative expansion:
\begin{align}
  \begin{split}
    \cfrac{\kappa}{2} \, \sqrt{-g} \,g^{\mu\nu} \, h_{\mu\nu}  &=\!\cfrac{\kappa}{2}\!\sum\limits_{p_1=0}^\infty \kappa^{p_1} C^{\rho_1\sigma_1\cdots\rho_{p_1}\sigma_{p_1}}_{(p_1)}  h_{\rho_1\sigma_1}\cdots  h_{\rho_{p_1}\sigma_{p_1}} \!\! \sum\limits_{p_2=0}^\infty \! (-1)^{p_2} \kappa^{p_2} I^{\mu\nu\lambda_1\tau_1\cdots\lambda_{p_2}\tau_{p_2}}_{(p_2+1)} h_{\lambda_1\tau_1}\cdots h_{\lambda_{p_2}\tau_{p_2}}h_{\mu\nu}\\
    &=\sum\limits_{n=1}^\infty \sum\limits_{k=1}^n \, \cfrac{(-1)^{k-1}}{2} ~ \kappa^n\,I_{(k)}^{\rho_1\sigma_1\cdots\rho_k\sigma_k}\,C_{(n-k)}^{\rho_{k+1}\sigma_{k+1}\cdots\rho_n\sigma_n}\,h_{\rho_1\sigma_1}\cdots h_{\rho_n\sigma_n} \\
    &= \sum\limits_{n=1}^\infty \kappa^n \,\sum\limits_{k=1}^n \cfrac{(-1)^{k-1}}{2} ~ I_{(k)}^{\rho_1\sigma_1\cdots\rho_k\sigma_k}\,C_{(n-k)}^{\rho_{k+1}\sigma_{k+1}\cdots\rho_n\sigma_n}\,h_{\rho_1\sigma_1}\cdots h_{\rho_n\sigma_n} .
  \end{split}
\end{align}
Comparing these expressions, we obtain the desired recursive relation.

The relation \eqref{the_C_recursion} is recursive because it defines the plain $C$-tensor of $n$-the order through plain $C$-tensors of lower orders. It would be much preferable to obtain a relation that expresses $C_{(n)}$ through $C_{(n-1)}$ alone, but this is impossible. The formula \eqref{the_C_recursion} shows that all low orders must contribute to the plain $C$-tensor. Moreover, because the volume factor $\sqrt{-g}$ is a simple object, there are few relations that one can use to search for similar recursive relations. Therefore, it is safe to assume that no simpler formula exists.

In the same way, one can define tensors describing the perturbative structure of the other factors involving $\sqrt{-g}$. Plain $C_{(1)}$, $C_{(2)}$, $C_{(3)}$, and any other $C_{(n)}$-tensors are defined as follows:
\begin{align}
  \begin{split}
    \sqrt{-g} \, g^{\mu\nu} \overset{\text{def}}{=}& \sum\limits_{n=0}^\infty \, \kappa^n \,C_{(1; n)}^{\mu\nu,\rho_1\sigma_1\cdots\rho_n\sigma_n} h_{\rho_1\sigma_1} \cdots h_{\rho_n\sigma_n} \, , \\
    \sqrt{-g} \, g^{\mu\nu}\,g^{\alpha\beta} \overset{\text{def}}{=}& \sum\limits_{n=0}^\infty \, \kappa^n \,C_{(2; n)}^{\mu\nu\alpha\beta,\rho_1\sigma_1\cdots\rho_n\sigma_n} h_{\rho_1\sigma_1} \cdots h_{\rho_n\sigma_n} \, , \\
    \sqrt{-g} \, g^{\mu\nu}\,g^{\alpha\beta}\,g^{\rho\sigma} \overset{\text{def}}{=}& \sum\limits_{n=0}^\infty \, \kappa^n \,C_{(3; n)}^{\mu\nu\alpha\beta\rho\sigma,\rho_1\sigma_1\cdots\rho_n\sigma_n} h_{\rho_1\sigma_1} \cdots h_{\rho_n\sigma_n} \, , \\
    \sqrt{-g} \, g^{\mu_1\nu_1}\cdots g^{\mu_l\nu_l} \overset{\text{def}}{=}& \sum\limits_{n=0}^\infty \, \kappa^n \,C_{ (l; n) }^{\mu_1\nu_1\cdots\mu_l\nu_l,\rho_1\sigma_1\cdots\rho_n\sigma_n} h_{\rho_1\sigma_1} \cdots h_{\rho_n\sigma_n} \, .
  \end{split}
\end{align}
The corresponding symmetric tensors are defined as follows.
\begin{align}\label{the_C_l_n_tensor_definition}
  \mathcal{C}_{(l; n)}^{\mu_1\nu_1\cdots\mu_l\nu_l,\rho_1\sigma_1\cdots\rho_n\sigma_n} \overset{\text{def}}{=} \cfrac{1}{2^n} \, \cfrac{1}{n!} \, \left[ C_{(l; n)}^{\mu_1\nu_1\cdots\mu_l\nu_l,\rho_1\sigma_1\cdots\rho_n\sigma_n} + \text{permutations} \right] .
\end{align}
Permutations account for all terms that make the $\mathcal{C}$-tensor symmetric with respect to permutations of indices within each index pair $\rho_i \leftrightarrow \sigma_i$ and with respect to permutations of any two index pairs $\{\rho_i,\sigma_i\} \leftrightarrow \{\rho_j,\sigma_j\}$. Let us highlight that the symmetrisation is performed only for indices contracted with the perturbation indices.

These definitions are also non-constructive, but it is possible to obtain the following recursive relation that defines them completely:
\begin{align}\label{the_C_l_n_recursion}
  C_{ (l; n) }^{\mu_1\nu_1\mu_2\nu_2 \cdots \mu_l\nu_l, \rho_1\sigma_1\cdots\rho_n\sigma_n} = \sum\limits_{p=0}^n (-1)^p \,I_{(1+p)}^{\mu_1\nu_1\rho_1\sigma_1\cdots\rho_p\sigma_p} C_{(l-1; n-p) }^{\mu_2\nu_2 \cdots \mu_l\nu_l, \rho_{p+1}\sigma_{p+1}\cdots\rho_n\sigma_n} .
\end{align}
This relation is not based on scaling but on a simple series multiplication. On the one hand, one can use the $C_{(l)}$ tensor definition:  
\begin{align}
  \sqrt{-g} \, g^{\mu_1\nu_1}\cdots g^{\mu_l\nu_l} \overset{\text{def}}{=}& \sum\limits_{n=0}^\infty \, \kappa^n \,C_{ (l; n) }^{\mu_1\nu_1\cdots\mu_l\nu_l,\rho_1\sigma_1\cdots\rho_n\sigma_n} h_{\rho_1\sigma_1} \cdots h_{\rho_n\sigma_n} \, .
\end{align}
On the other hand, one can separate one metric and expand it separately:
\begin{align}
  \begin{split}
    & \sqrt{-g} \, g^{\mu_1\nu_1}\, g^{\mu_2\nu_2}\cdots g^{\mu_l\nu_l} = g^{\mu_1\nu_1} \sqrt{-g} \, g^{\mu_2\nu_2}\cdots g^{\mu_l\nu_l}\\
    &= \sum\limits_{n_1=0}^\infty (-1)^{n_1} \kappa^{n_1} I_{(n_1+1)}^{\mu_1\nu_1\rho_1\sigma_1\cdots\rho_{n_1}\sigma_{n_1}} h_{\rho_1\sigma_1}\cdots h_{\rho_{n_1}\sigma_{n_1}}  \sum\limits_{n_2=0}^\infty \kappa^{n_2} C_{ (l-1; n_2)}^{\mu_2\nu_2\cdots\mu_l\nu_l,\lambda_1\tau_1\cdots\lambda_{n_2}\tau_{n_2} } h_{\lambda_1\tau_1} \cdots h_{\lambda_{n_2}\tau_{n_2}}\\
    &=\sum\limits_{n_1=0}^\infty \sum\limits_{n_2=0}^\infty (-1)^{n_1} \kappa^{n_1 + n_2} I_{(n_1+1)}^{\mu_1\nu_1\rho_1\sigma_1\cdots\rho_{n_1}\sigma_{n_1}} C_{ (l-1; n_2)}^{\mu_2\nu_2\cdots\mu_l\nu_l,\rho_{n_1+1}\sigma_{n_1+1}\cdots\rho_{n_1+n_2}\sigma_{n_1+n_2}} h_{\rho_1\sigma_1}\cdots h_{\rho_{n_1+n_2}\sigma_{n_1+n_2}}.
  \end{split}
\end{align}
One shall change the summation order in this expression to obtain the discussed expression:
\begin{align}
  \begin{split}
    & \sqrt{-g} \, g^{\mu_1\nu_1}\, g^{\mu_2\nu_2}\cdots g^{\mu_l\nu_l} = \\
    &=\sum\limits_{n=0}^\infty \sum\limits_{p=0}^n \kappa^n (-1)^p  I_{(1 + p)}^{\mu_1\nu_1\rho_1\sigma_1\cdots\rho_p \sigma_p } C_{ (l-1; n_2)}^{\mu_2\nu_2\cdots\mu_l\nu_l,\rho_{p+1}\sigma_{p+1}\cdots\rho_n \sigma_n } h_{\rho_1\sigma_1}\cdots h_{\rho_n \sigma_n }.
  \end{split}
\end{align}

Similar definitions and recursive relations also hold for factors involving the vierbein $\mathfrak{e}_\mu{}^\nu$ \cite{Prinz:2020nru}:
\begin{align}
  \begin{split}
    \mathfrak{e}_\mu {}^\nu \overset{\text{def}}{=}& \sum\limits_{n=0}^\infty \kappa^n E_{\mu~(1+n)}^{~~\nu\rho_1\sigma_1\cdots\rho_n\sigma_n} h_{\rho_1\sigma_1}\cdots h_{\rho_n\sigma_n} \\
    =& \sum\limits_{n=0}^\infty \, \kappa^n \binom{-\frac12 }{~n} I_{\mu~(1+n)}^{~~\nu\rho_1\sigma_1\cdots\rho_n\sigma_n} h_{\rho_1\sigma_1} \cdots h_{\rho_n\sigma_n} \, ,\\
    \sqrt{-g} \, \mathfrak{e}_\mu {}^\nu \overset{\text{def}}{=}& \sum\limits_{n=0}^\infty \kappa^n {C_E}_{\mu ~(n)}^{~\nu,\rho_1\sigma_1\cdots\rho_n\sigma_n} h_{\rho_1\sigma_1}\cdots h_{\rho_n\sigma_n} \\
    =&\sum\limits_{n=0}^\infty \, \kappa^n ~\sum\limits_{p=0}^n  \binom{-\frac12}{~n} C_{(p)}^{\rho_1\sigma_1\cdots\rho_p\sigma_p} I_{\mu~(1+n-p)}^{~~\nu\rho_{p+1}\sigma_{p+1}\cdots\rho_n\sigma_n}  h_{\rho_1\sigma_1} \cdots h_{\rho_n\sigma_n} \, .
  \end{split}
\end{align}
Here $\binom{x}{y}$ is the binomial coefficient. Only $C_E$ tensor can enter a Lagrangian, and its symmetric counterpart reads
\begin{align}\label{the_C_E_recursion}
  { \mathcal{C}_E }_{\mu~(n)}^{~~\nu,\rho_1\sigma_1\cdots\rho_n\sigma_n} \overset{\text{def}}{=} \cfrac{1}{2^n} \, \cfrac{1}{n!} \, \left[ {C_E}_{\mu~(n)}^{~\nu,\rho_1\sigma_1\cdots\rho_n\sigma_n} + \text{permutations} \right].
\end{align}
Permutations account for all terms that make the $\mathcal{C}_E$-tensor symmetric with respect to permutations of indices within each index pair $\rho_i\leftrightarrow\sigma_i$, and with respect to permutations of any two index pairs $\{\rho_i,\sigma_i\} \leftrightarrow \{\rho_j,\sigma_j\}$.

A similar recurrent relation holds for the $C_E$ tensor and its proof is similar to the $C_{(l)}$ case:
\begin{align}
  {C_E}_{\mu~(n)}^{~\nu, \rho_1\sigma_1\cdots\rho_n\sigma_n} = \sum\limits_{p=0}^n E_{\mu~(1+p)}^{~~\nu\rho_1\sigma_1\cdots\rho_p\sigma_p} C_{(n-p)}^{\rho_{p+1}\sigma_{p+1}\cdots\rho_n\sigma_n} .
\end{align}

Relations \eqref{the_C_recursion}, \eqref{the_C_l_n_recursion}, and \eqref{the_C_E_recursion} are the main results of this Section. They provide an alternative way to calculate the perturbative structure of the corresponding objects. They improve computational efficiency, making them essential for the higher-order perturbation theory. The new version of \texttt{FeynGrav} uses these formulae.

Lastly, similar relations hold for an arbitrary metric. One shall only replace the flat metric $\eta_{\mu\nu}$ with the arbitrary background $\overline{g}_{\mu\nu}$ in the definition of all tensors. However, in that more sophisticated case, the recurrent relation may not improve computational efficiency that strongly. Since \texttt{FeynGrav} operates with \texttt{FeynCalc}, one does not operate with metric components but with the particular objects defined within \texttt{FeynCalc}. To put it otherwise, \texttt{FeynCalc} does not directly refer to the metric component but operates with object index contraction. The discussed recursive relations involve fewer operations, which is the reason behind their efficiency. The same logic may not hold for perturbations about an arbitrary metric, so the efficiency is not guaranteed.

\section{Horndeski gravity}\label{Section_Horndeski}

The Horndeski theory is the most general scalar-tensor theory without coupling between the scalar field and the regular matter that admits second-order field equations. The theory was discovered in \cite{Horndeski:1974wa} and independently rediscovered in \cite{Kobayashi:2011nu}. The second differential order of field equations ensures the theory is free from the Ostrogradsky instability \cite{Ostrogradsky:1850fid,Woodard:2015zca,Woodard:2006nt}.

The theory is constituted by the four following Lagrangians:
\begin{align}
  \mathcal{A} = \int d^4 x \sqrt{-g} \left[ \mathcal{L}_2 + \mathcal{L}_3 + \mathcal{L}_4 + \mathcal{L}_5 \right] ,
\end{align}
\begin{align*}
  \mathcal{L}_2 &= G_2 (\phi,X) \,,\\
  \mathcal{L}_3 &= G_3 (\phi,X) \, \square \phi \,,\\
  \mathcal{L}_4 &= G_4 (\phi,X) \, R + G_{4,X} \left[ \left(\square\phi\right)^2  - \left(\nabla_\mu\nabla_\nu\phi\right)^2 \right] ,\\
  \mathcal{L}_5 &= G_5 (\phi,X) \, G_{\mu\nu} \nabla^\mu\nabla^\nu\phi - \cfrac16\, G_{5,X} \left[ \left( \square\phi\right)^3 - 3 \, \square\phi \, \left(\nabla_\mu\nabla_\nu\phi\right)^2 + 2 \left(\nabla_\mu\nabla_\nu \phi\right)^3 \right] .
\end{align*}
Here $G_i = G_i (\phi,X)$ are functions of the scalar field $\phi$ and its kinetic term $X = g^{\mu\nu} \,\nabla_\mu \phi \, \nabla_\nu\phi$, while $G_{4,X}$ and $G_{5,X}$ note derivatives with respect to the kinetic term.

The theory shows that there are only four couplings with the desired features. Interactions of the $G_2$ class correspond to the minimal coupling to the gravity of non-minimally interacting scalar fields. The coupling is minimal since it does not involve Christoffel symbols. Interaction of the $G_3$ class corresponds to the simplest non-minimal coupling to gravity. The presence of a Christoffel symbol shows that the scalar field couples to a graviton momentum. The coupling is non-minimal since it involves a single Christoffel symbol, but it is minimal because only one symbol is present. The Christoffel symbol can be removed if $G_{3, X}=0$, but in that case, the interaction becomes redundant as it reduces to the $G_2$ case with an integration by parts. Interactions of $G_4$ and $G_5$ classes involve up to two and three Christoffel symbols.

Counting the Christoffel symbols shows that the Horndeski theory can be understood as a theory of coupling between a scalar field and graviton momenta. A healthy coupling cannot involve more than three graviton momenta. In turn, such coupling is strongly constrained, and only the Horndeski interactions are allowed.

These considerations make the Horndeski theory an essential part of perturbative quantum gravity. This paper considers all the Horndeski interactions and presents explicit formulae for the corresponding perturbative expansions. Since the interactions involve many derivatives, the formulae occupy much space and are computationally challenging. Consequently, despite having all the required analytic expressions, we implement only a few selected cases in the new version of \texttt{FeynGrav}, as discussed in Section \ref{Section_V3}. It shall also be noted that the $G_5$ interaction class naturally involves the standard Einstein-Hilbert term describing pure general relativity. For the sake of simplicity, we separate this term and discuss only contributions describing interactions. The theory was already briefly discussed in \cite{Latosh:2024anf}. This paper addresses it in more detail and also discusses its implementations.

\subsection{Horndeski $G_2(\phi,X)$ class}

The interactions of the $G_2$ class of Horndeski theory describe the minimal coupling between the scalar field and gravity. The coupling is minimal because it does not involve graviton momenta. The corresponding Feynman rules are obtained as follows.

First and foremost, the function $G_2$ shall be smooth enough to admit a power series expansions:
\begin{align}
  \begin{split}
    G_2(\phi,X) & = \sum\limits_{a=0}^\infty \sum\limits_{b=0}^\infty \lambda_{(a,b)} \, \phi^a \, X^b \\
    & = \sum\limits_{a=0}^\infty \sum\limits_{b=0}^\infty \lambda_{(a,b)} \, \phi^a \, g^{\alpha_1\beta_1} \cdots g^{\alpha_b\beta_b} \pd_{\alpha_1}\phi \, \pd_{\beta_1} \phi \cdots \pd_{\alpha_b} \phi \, \pd_{\beta_b} \phi .
  \end{split}
\end{align}
Here $\lambda_{(a,b)}$ is a dimensional coupling with the mass dimension $ 4 (1-b) - a $. Without the loss of generality, one can consider a single term of this expansion:
\begin{align}\label{HorndeskiG2_Parametrisation}
  \int d^4 x \sqrt{-g} \, G_2(\phi, X) \to \int d^4 x \sqrt{-g} \, \lambda_{(a,b)} \, \phi^a \, X^b \,.
\end{align}

Secondly, one shall use the standard Fourier transformations to describe the momentum structure of the interaction.
\begin{align}
  \begin{split}
    &\int d^4 x \sqrt{-g} \, \lambda_{(a,b)} \, \phi^a \, X^ b = \int d^4 x \sqrt{-g} \, g^{\alpha_1\beta_1} \cdots g^{\alpha_b \beta_b}  \,\phi^a \, \pd_{\alpha_1} \phi \, \pd_{\beta_1} \phi \cdots \pd_{\alpha_b} \phi \pd_{\beta_b} \phi  \\
    &=\sum\limits_{n=0}^\infty \int \prod\limits_{i=1}^n \cfrac{d^4 k_i}{(2\pi)^4} \, h_{\rho_i\sigma_i}(k_i) \, \prod\limits_{j=1}^{a + 2b} \cfrac{d^4 p_j}{(2\pi)^4} \, \phi(p_j) \, (2 \pi)^4 \delta \left( \sum k_i + \sum p_i \right)\\
    & \hspace{10pt} \times\kappa^n \, \lambda_{(a,b)} ~ \mathcal{C}_{ (b; n)}^{\alpha_1\beta_1\cdots\alpha_b\beta_b , \rho_1\sigma_1\cdots \rho_n\sigma_n} (p_1)_{\alpha_1} (p_2)_{\beta_1} \cdots (p_{2b-1})_{\alpha_b} (p_{2b})_{\beta_b} \,.
  \end{split}
\end{align}

Many formulae in this paper rely on the Fourier transformation. We do not discuss their derivation in detail since they are only required to perform the transformation correctly. Similar formulae are discussed in detail in previous publications \cite{Latosh:2022ydd,Latosh:2023zsi}.

Lastly, one shall construct the rule corresponding Feynman rules:
\begin{align}
  \nonumber \\
  \begin{gathered}
    \begin{fmffile}{Diags/G2_vertex}
      \begin{fmfgraph*}(40,40)
        \fmfleft{L1,L2}
        \fmfright{R1,R2}
        \fmf{dbl_wiggly}{L1,V}
        \fmf{dbl_wiggly}{L2,V}
        \fmf{dashes}{V,R1}
        \fmf{dashes}{V,R2}
        \fmfdot{V}
        \fmffreeze
        \fmf{dots}{L1,L2}
        \fmf{dots}{R1,R2}
        \fmflabel{$\rho_1\sigma_1,k_1$}{L1}
        \fmflabel{$\rho_n\sigma_n,k_n$}{L2}
        \fmflabel{$p_1$}{R1}
        \fmflabel{$p_{a+2b}$}{R2}
        \fmflabel{$\lambda_{(a,b)}$}{V}
        \fmfv{$\lambda_{(a,b)}$,label.angle=90,label.dist=15pt}{V}
      \end{fmfgraph*}
    \end{fmffile}
  \end{gathered}
  \hspace{20pt}
  \begin{split}
    & = i\, \kappa^n \,(-1)^b\,\lambda_{(a,b)} \, \mathcal{C}_{(b; n)}^{\alpha_1\beta_1\cdots\alpha_b\beta_b , \rho_1\sigma_1\cdots \rho_n\sigma_n} (p_1)_{\alpha_1} (p_2)_{\beta_1} \cdots (p_{2b-1})_{\alpha_b} (p_{2b})_{\beta_b} \\
    & \hspace{10pt} + \text{permutations} .
  \end{split} \\ \nonumber
\end{align}
In this expression, the permutation term accounts for all possible permutations of $a + 2 \, b$ scalars and $n$ gravitons. When $b=0$, the interaction reduces to the standard scalar field potential, considered in \cite{Latosh:2023zsi}.

\subsection{Horndeski $G_3(\phi,X)$ class}

Interactions of $G_3$ class describe the simplest non-minimal coupling between the scalar field and gravity. The interaction involves a single graviton momentum due to the presence of the Christoffel symbol. The presence of the Christoffel symbol can be excluded using integration by parts in an exceptional case $G_{3, X}=0$. However, in that case, the interaction is also reduced to the $G_2$ class, making this case redundant. Because of this reason, we implicitly assume $G_{3, X}\not =0$ within this section.

We shall separate the Christoffel symbols explicitly to proceed with deriving interaction rules.
\begin{align}
  \begin{split}
    \int d^4 x \sqrt{-g} \, G_3 \, \square \phi = \int d^4 x \sqrt{-g} \, g^{\mu\nu} G_3 \,\pd_\mu \pd_\nu \phi -\int d^4 x \sqrt{-g} \, g^{\mu\nu} g^{\alpha\beta}\,\Gamma_{\alpha\mu\nu} G_3 \, \pd_\beta \phi  .
  \end{split}
\end{align}

Similar to the previous case, $G_3$ should be smooth enough to admit a power series expansion:
\begin{align}
  G_3(\phi,X) = \sum\limits_{a=1}^\infty \sum\limits_{b=1}^\infty \, \Theta_{(a,b)} \, \phi^a \, X^b .
\end{align}
One can study a single term of the expansion without the loss of generality:
\begin{align}
  \int d^4 x \sqrt{-g}\, G_3 (\phi,X) \, \square \phi \to \int d^4 x \sqrt{-g} \, \Theta_{(a,b)} \, \phi^a \, X^b \, \square \phi .
\end{align}
Here, $\Theta_{(a,b)}$ is a dimensional coupling with the mass dimension $4(1-b)-a-3$.

After this simplification, one can make the perturbative structure of the interaction explicit.
\begin{align}
  \begin{split}
    &\int d^4 x \sqrt{-g} \, \Theta_{(a,b)}\, \phi^a\, X^b\, \square \phi \\
    & = \int d^4 x \sqrt{-g} \, g^{\mu\nu} \Theta_{(a,b)} \,\phi^a \, X^b \, \pd_\mu \pd_\nu \phi - \int d^4 x \sqrt{-g} \, g^{\mu\nu} g^{\rho\sigma} \Gamma_{\rho\mu\nu} \Theta_{(a,b)} \,\phi^a \, X^b \, \pd_\sigma \phi \\
    & = \sum\limits_{n=0}^\infty \int \prod\limits_{i=1}^n \cfrac{d^4 k_i}{(2\pi)^4}\,h_{\rho_i\sigma_i}(k_i) \!\!\!\! \prod\limits_{j=1}^{a+2 b+1} \!\! \cfrac{d^4 p_j}{(2\pi)^4}\, \phi(p_j) ~(2\pi)^4 \delta\left( \sum k_i + \sum p_i \right)\\
    &\hspace{10pt}\times\kappa^n (-1)^{b+1} \Theta_{(a,b)}\mathcal{C}_{(1+b;n)}^{\mu\nu\alpha_1\beta_1\cdots\alpha_b\beta_b,\rho_1\sigma_1\cdots\rho_n\sigma_n} (p_1)_{\alpha_1}(p_2)_{\beta_1}\cdots (p_{2b-1})_{\alpha_b}(p_{2b})_{\beta_b} (p_{2b+1})_\mu (p_{2b+1})_\nu\\
    & \hspace{10pt}-\sum\limits_{n=1}^\infty \int \prod\limits_{i=1}^n \cfrac{d^4 k_i}{(2\pi)^4}\,h_{\rho_i\sigma_i}(k_i) \!\!\!\! \prod\limits_{j=1}^{a+2 b+1} \!\! \cfrac{d^4 p_j}{(2\pi)^4} \, \phi(p_j) ~(2\pi)^4 \delta\left( \sum k_i + \sum p_i \right)\\
    &\hspace{20pt}\times\kappa^n(-1)^{b+1} \Theta_{(a,b)} \mathcal{C}_{(2+b; n-1)}^{\mu\nu\rho\sigma\alpha_1\cdots\beta_b,\rho_1\cdots\sigma_{n-1}} (\Gamma_{\rho\mu\nu})^{\lambda\rho_n\sigma_n} (k_n)_\lambda (p_1)_{\alpha_1}(p_2)_{\beta_1}\cdots (p_{2b-1})_{\alpha_b}(p_{2b})_{\beta_b} (p_{2b+1})_\sigma .
  \end{split}
\end{align}

The first part of this expression contributes at $\kappa^0$ level and describes the self-interaction of the scalar field. The second part does not contribute at $\kappa^0$ level and only describes the coupling between scalars and gravitons. Similar features are typical for coupling involving a few Christoffel symbols. Such interaction terms do not contribute to low orders of $\kappa$. We highlight this feature wherever it occurs. We also include the terms in all formulas to make them more compact.

Finally, we obtain the interaction rule by appropriately symmetrising the expression that we have obtained.
\begin{align}
  \nonumber \\
  \begin{gathered}
    \begin{fmffile}{Diags/G3_vertex}
      \begin{fmfgraph*}(40,40)
        \fmfleft{L1,L2}
        \fmfright{R1,R2}
        \fmf{dbl_wiggly}{L1,V}
        \fmf{dbl_wiggly}{L2,V}
        \fmf{dashes}{V,R1}
        \fmf{dashes}{V,R2}
        \fmfdot{V}
        \fmffreeze
        \fmf{dots}{L1,L2}
        \fmf{dots}{R1,R2}
        \fmflabel{$\rho_1\sigma_1,k_1$}{L1}
        \fmflabel{$\rho_n\sigma_n,k_n$}{L2}
        \fmflabel{$p_1$}{R1}
        \fmflabel{$p_{a+2b+1}$}{R2}
        \fmflabel{$\Theta_{(a,b)}$}{V}
        \fmfv{$\Theta_{(a,b)}$,label.angle=90,label.dist=15pt}{V}
      \end{fmfgraph*}
    \end{fmffile}
  \end{gathered}
  \hspace{40pt}
  \begin{split}
    =& i \,\kappa^n (-1)^{b+1} \Theta_{(a,b)} \Bigg[ \mathcal{C}_{(1+b;n)}^{\mu\nu\alpha_1\beta_1\cdots\alpha_b\beta_b,\rho_1\sigma_1\cdots\rho_n\sigma_n} (p_{2b+1})_\mu (p_{2b+1})_\nu \\
      & \hspace{20pt} - \mathcal{C}_{(2+b; n-1)}^{\mu\nu\rho\sigma\alpha_1\cdots\beta_b,\rho_1\sigma_1\cdots\sigma_{n-1}\rho_{n-1}} (\Gamma_{\rho\mu\nu})^{\lambda\rho_n\sigma_n} (k_n)_\lambda (p_{2b+1})_\sigma \Bigg] \\
    & \times (p_1)_{\alpha_1}(p_2)_{\beta_1}\cdots (p_{2b-1})_{\alpha_b}(p_{2b})_{\beta_b} + \text{permutations} .
  \end{split}
\end{align}
The permutation term includes all possible permutations of $2\, b +a +1$ scalar fields and $n$ gravitons. Let us highlight again that only the term in this expression contributes at $\kappa^0$ level.
  
\subsection{Horndeski $G_4(\phi,X)$ class}

Interactions of $G_4$ class take a special place within the Horndeski theory since pure general relativity belongs to this class. The pure Einstein-Hilbert term was considered in previous publications \cite{Latosh:2022ydd,Latosh:2023zsi}, so we will not consider it in this section.

The exclude the Einstein-Hilbert term, we assume that the $G_4$ function admits the following power series expansion:
\begin{align}
  G_4 = -\cfrac{2}{\kappa^2} + \sum\limits_{a,b} \Upsilon_{(a,b)} \phi^a \, X^b .
\end{align}
Here $\Upsilon_{(a,b)}$ is a dimensional coupling with the mass dimension $4 (1-b) - a - 2$. To account only for the interaction terms, we study only the following term without the loss of generality:
\begin{align}
  G_4 \to \Upsilon_{(a,b)} \phi^a \,X^b.
\end{align}

Further, we shall make the Christoffel symbols explicit with the following formula:
\begin{align}
  R = g^{\mu\nu} g^{\alpha\beta} \pd_\mu \left[ \Gamma_{\nu\alpha\beta} - \Gamma_{\alpha\beta\nu} \right] + g^{\mu\nu} g^{\alpha\beta} g^{\rho\sigma} \left[ \Gamma_{\mu\alpha\rho} \Gamma_{\nu\beta\sigma} - \Gamma_{\mu\alpha\beta} \Gamma_{\nu\rho\sigma} \right].
\end{align}
In turn, this allows us to obtain the following explicit expression for the interaction Lagrangian:
\begin{align}
  \begin{split}
    & \int d^4 x \sqrt{-g} \, \Upsilon_{(a,b)}\,\phi^a \, X^{b-1} \left[ X R \!+\! b \left( \! \left(\square\phi\right)^2 \!-\! \left(\nabla_\mu \nabla_\nu\phi\right)^2\right) \right]\\
    &= \Upsilon_{(a,b)} \int d^4 x \Bigg[ \sqrt{-g} \,g^{\mu\nu} g^{\alpha\beta} \pd_\mu \left[ \Gamma_{\nu\alpha\beta} - \Gamma_{\alpha\beta\nu} \right] \phi^a X^b + \sqrt{-g} \,g^{\mu\nu} g^{\alpha\beta} g^{\rho\sigma} \left[ \Gamma_{\mu\alpha\rho} \Gamma_{\nu\sigma\beta} - \Gamma_{\mu\alpha\beta} \Gamma_{\nu\rho\sigma} \right] \phi^a X^b \\
      &\hspace{70pt} + b \sqrt{-g} \left( g^{\mu\nu} g^{\alpha\beta} - g^{\mu\alpha} g^{\nu\beta} \right) \nabla_\mu\nabla_\nu\phi \nabla_\alpha\nabla_\beta \phi \,\phi^a X^{b-1} \Bigg] \\
    &= \Upsilon_{(a,b)} \int d^4 x \Bigg[ \sqrt{-g} \,g^{\mu\nu} g^{\alpha\beta} \pd_\mu \left[ \Gamma_{\nu\alpha\beta} - \Gamma_{\alpha\beta\nu} \right] \phi^a X^b + \sqrt{-g} \,g^{\mu\nu} g^{\alpha\beta} g^{\rho\sigma} \left[ \Gamma_{\mu\alpha\rho} \Gamma_{\nu\sigma\beta} - \Gamma_{\mu\alpha\beta} \Gamma_{\nu\rho\sigma} \right] \phi^a X^b \\
      &\hspace{70pt} + b \sqrt{-g} \left( g^{\mu\nu} g^{\alpha\beta} - g^{\mu\alpha} g^{\nu\beta} \right) \pd_\mu\pd_\nu\phi \, \pd_\alpha\pd_\beta \phi \,\phi^a X^{b-1} \\
      &\hspace{70pt} - 2\, b \sqrt{-g} \left( g^{\mu\nu} g^{\alpha\beta} - g^{\mu\alpha} g^{\nu\beta} \right) g^{\rho\sigma} \Gamma_{\rho\mu\nu}\, \pd_\sigma\phi \, \pd_\alpha\pd_\beta \phi \,\phi^a X^{b-1} \\
      &\hspace{70pt} + b \sqrt{-g} \left( g^{\mu\nu} g^{\alpha\beta} - g^{\mu\alpha} g^{\nu\beta} \right) g^{\rho\sigma} g^{\lambda\tau} \Gamma_{\rho\mu\nu} \Gamma_{\lambda\alpha\beta} \pd_\sigma\phi \, \pd_\tau \phi \,\phi^a X^{b-1} \Bigg].
  \end{split}
\end{align}

Lastly, one can make the perturbative structure of this interaction explicit.
\begin{align}
  \begin{split}
    &\int d^4 x \sqrt{-g} \Upsilon_{(a,b)}\,\phi^a \, X^{b-1} \left[ X R \!+\! b \left( \! \left(\square\phi\right)^2 \!-\! \left(\nabla_\mu \nabla_\nu\phi\right)^2\right) \right] \\
    &=\sum\limits_{n=1}^\infty\int \prod\limits_{i=1}^n \cfrac{d^4 k_i}{(2\pi)^4} \, h_{\rho_i\sigma_i}(k_i) \prod\limits_{j=1}^{a+2b} \cfrac{d^4 p_j}{(2\pi)^4} \, \phi(p_j) \,(2\pi)^4 \delta\left( \sum k_i + \sum p_j \right)\\
    & \hspace{8pt} \times \kappa^n \Upsilon_{(a,b)} (-1)^{b+1} \mathcal{C}_{(2+b,n-1)}^{\mu\nu\alpha\beta\alpha_1\beta_1\cdots\alpha_b\beta_b,\rho_1\cdots\sigma_{n\!-\!1}} (k_n)_\mu (k_n)_\lambda \left[\left(\Gamma_{\nu\alpha\beta}\right)^{\lambda\rho_n\sigma_n} - \left(\Gamma_{\alpha\beta\nu}\right)^{\lambda\rho_n\sigma_n}\right] (p_1)_{\alpha_1}\cdots(p_{2b})_{\beta_b}
  \end{split}
\end{align}
\begin{align*}
  &+\sum\limits_{n=2}^\infty\int \prod\limits_{i=1}^n \cfrac{d^4 k_i}{(2\pi)^4} \, h_{\rho_i\sigma_i}(k_i) \prod\limits_{j=1}^{a+2b} \cfrac{d^4 p_j}{(2\pi)^4} \, \phi(p_j) \, (2\pi)^4 \delta\left( \sum k_i + \sum p_j \right)\\
  &\hspace{8pt} \times \kappa^n \Upsilon_{(a,b)} (-1)^{b+1} \mathcal{C}_{(3+b,n-2)}^{\mu\nu\alpha\beta\rho\sigma\alpha_1\cdots\beta_b,\rho_1\cdots\sigma_{n\!-\!2}} (k_{n\!-\!1})_{\lambda_1} (k_n)_{\lambda_2} \\
  & \hspace{16pt} \times\left[ \left(\Gamma_{\mu\alpha\rho}\right)^{\lambda_1\rho_{n\!-\!1}\sigma_{n\!-\!1}}\left(\Gamma_{\nu\sigma\beta}\right)^{\lambda_2\rho_n\sigma_n} - \left(\Gamma_{\mu\alpha\beta}\right)^{\lambda_1\rho_{n\!-\!1}\sigma_{n\!-\!1}}\left(\Gamma_{\nu\rho\sigma}\right)^{\lambda_2\rho_n\sigma_n} \right] (p_1)_{\alpha_1}\cdots(p_{2b})_{\beta_b} \\
  & +\sum\limits_{n=0}^\infty\int \prod\limits_{i=1}^n \cfrac{d^4 k_i}{(2\pi)^4} \, h_{\rho_i\sigma_i}(k_i) \prod\limits_{j=1}^{a+2b} \cfrac{d^4 p_j}{(2\pi)^4} \, \phi(p_j) \, (2\pi)^4 \delta\left( \sum k_i + \sum p_j \right)\\
  & \hspace{8pt} \times \kappa^n \Upsilon_{(a,b)} \,b\, (-1)^{b+1} \left[ \mathcal{C}_{(1+b,n)}^{\mu\nu\alpha\beta\alpha_1\cdots\beta_{b-1},\rho_1\cdots\sigma_n} -\mathcal{C}_{(1+b,n)}^{\mu\alpha\nu\beta\alpha_1\cdots\beta_{b-1},\rho_1\cdots\sigma_n} \right] \\
  &\hspace{16pt} \times (p_1)_{\alpha_1} (p_2)_{\beta_1}\cdots (p_{2b-3})_{\alpha_{b-1}} (p_{2b-2})_{\beta_{b-1}} (p_{2b-1})_\mu (p_{2b-1})_\nu (p_{2b})_\alpha (p_{2b})_\beta \\
  &-\sum\limits_{n=1}^\infty\int \prod\limits_{i=1}^n \cfrac{d^4 k_i}{(2\pi)^4} \, h_{\rho_i\sigma_i}(k_i) \prod\limits_{j=1}^{a+2b} \cfrac{d^4 p_j}{(2\pi)^4} \, \phi(p_j) \, (2\pi)^4 \delta\left( \sum k_i + \sum p_j \right)\\
  &\hspace{8pt} \times \kappa^n \Upsilon_{(a,b)} \, 2\, b\, (-1)^{b+1} \left[ \mathcal{C}_{(2+b,n-1)}^{\mu\nu\alpha\beta\rho\sigma\alpha_1\cdots\beta_{b-1},\rho_1\cdots\sigma_{n\!-\!1}} -\mathcal{C}_{(2+b,n-1)}^{\mu\alpha\nu\beta\rho\sigma\alpha_1\cdots\beta_{b-1},\rho_1\cdots\sigma_{n\!-\!1}} \right] (k_n)_\lambda \left( \Gamma_{\rho\mu\nu} \right)^{\lambda\rho_n\sigma_n}\\
  &\hspace{16pt} \times (p_1)_{\alpha_1}\cdots(p_{2b-2})_{\beta_{b-1}} (p_{2b-1})_\sigma (p_{2b})_\alpha (p_{2b})_\beta \\
  &+\sum\limits_{n=2}^\infty\int \prod\limits_{i=1}^n \cfrac{d^4 k_i}{(2\pi)^4} \, h_{\rho_i\sigma_i}(k_i) \prod\limits_{j=1}^{a+2b} \cfrac{d^4 p_j}{(2\pi)^4} \, \phi(p_j) \, (2\pi)^4 \delta\left( \sum k_i + \sum p_j \right)\\
  &\hspace{8pt} \times \kappa^n \Upsilon_{(a,b)} \, b\, (-1)^{b+1} \left[ \mathcal{C}_{(3+b,n-2)}^{\mu\nu\alpha\beta\rho\sigma\lambda\tau\alpha_1\cdots\beta_{b-1},\rho_1\cdots\sigma_{n\!-\!2}} -\mathcal{C}_{(3+b,n-2)}^{\mu\alpha\nu\beta\rho\sigma\lambda\tau\alpha_1\cdots\beta_{b-1},\rho_1\cdots\sigma_{n\!-\!2}} \right] \left( \Gamma_{\rho\mu\nu} \right)^{\lambda_1\rho_{n\!-\!1}\sigma_{n\!-\!1}}\left( \Gamma_{\lambda\alpha\beta} \right)^{\lambda_2\rho_n\sigma_n}\\
  &\hspace{16pt} \times (k_{n\!-\!1})_{\lambda_1} (k_n)_{\lambda_2}  (p_1)_{\alpha_1}\cdots(p_{2b-2})_{\beta_{b-1}} (p_{2b-1})_\sigma (p_{2b})_\tau .
\end{align*}
Only the third term contributes at $\kappa^0$ level in this expression. The first and fourth terms contribute at the $\kappa^1$ level, and the second and fifth terms contribute at the $\kappa^2$ level. Equivalently, only the second and fifth terms describe the coupling of a scalar with two graviton momenta, while the other terms ensure that the interaction is healthy.

This expression corresponds to the following interaction rule
\begin{align}
  \nonumber \\
  \begin{split}
    & \hspace{40pt}
    \begin{gathered}
      \begin{fmffile}{Diags/G4_vertex}
        \begin{fmfgraph*}(40,40)
          \fmfleft{L1,L2}
          \fmfright{R1,R2}
          \fmf{dbl_wiggly}{L1,V}
          \fmf{dbl_wiggly}{L2,V}
          \fmf{dashes}{V,R1}
          \fmf{dashes}{V,R2}
          \fmfdot{V}
          \fmffreeze
          \fmf{dots}{L1,L2}
          \fmf{dots}{R1,R2}
          \fmflabel{$\rho_1\sigma_1,k_1$}{L1}
          \fmflabel{$\rho_n\sigma_n,k_n$}{L2}
          \fmflabel{$p_1$}{R1}
          \fmflabel{$p_{a+2b}$}{R2}
          \fmflabel{$\Upsilon_{a,b}$}{V}
          \fmfv{$\Upsilon_{a,b}$,label.angle=90,label.dist=15pt}{V}
        \end{fmfgraph*}
      \end{fmffile}
    \end{gathered}
    = i\, \kappa^n\, (-1)^{b+1} \Upsilon_{(a,b)} (p_1)_{\alpha_1} (p_2)_{\beta_1}\cdots (p_{2b-1})_{\alpha_b} (p_{2b})_{\beta_b} \\ \\
    & \times \!\!\Bigg[ \mathcal{C}_{(2+b,n-1)}^{\mu\nu\alpha\beta\alpha_1\beta_1\cdots\alpha_b\beta_b,\rho_1\cdots\sigma_{n\!-\!1}} (k_n)_\mu (k_n)_\lambda \left[\left(\Gamma_{\nu\alpha\beta}\right)^{\lambda\rho_n\sigma_n} - \left(\Gamma_{\alpha\beta\nu}\right)^{\lambda\rho_n\sigma_n}\right]\\
      & \hspace{8pt} +\mathcal{C}_{(3+b,n-2)}^{\mu\nu\alpha\beta\rho\sigma\alpha_1\cdots\beta_b,\rho_1\cdots\sigma_{n\!-\!2}} (k_{n\!-\!1})_{\lambda_1} (k_n)_{\lambda_2} \left[ \left(\Gamma_{\mu\alpha\rho}\right)^{\lambda_1\rho_{n\!-\!1}\sigma_{n\!-\!1}}\left(\Gamma_{\nu\sigma\beta}\right)^{\lambda_2\rho_n\sigma_n} \!\!-\! \left(\Gamma_{\mu\alpha\beta}\right)^{\lambda_1\rho_{n\!-\!1}\sigma_{n\!-\!1}}\left(\Gamma_{\nu\rho\sigma}\right)^{\lambda_2\rho_n\sigma_n} \right] \! \Bigg] \\
    & + i\, \kappa^n\, (-1)^{b+1} \Upsilon_{(a,b)}\, b\, (p_1)_{\alpha_1} (p_2)_{\beta_1}\cdots (p_{2b-3})_{\alpha_{b-1}} (p_{2b-2})_{\beta_{b-1}} \\
    & \hspace{8pt}\times\Bigg[ \left[ \mathcal{C}_{(1+b,n)}^{\mu\nu\alpha\beta\alpha_1\cdots\beta_{b-1},\rho_1\cdots\sigma_n} -\mathcal{C}_{(1+b,n)}^{\mu\alpha\nu\beta\alpha_1\cdots\beta_{b-1},\rho_1\cdots\sigma_n} \right] (p_{2b-1})_\mu (p_{2b-1})_\nu (p_{2b})_\alpha (p_{2b})_\beta \\
      & \hspace{16pt} - 2\, \left[ \mathcal{C}_{(2+b,n-1)}^{\mu\nu\alpha\beta\rho\sigma\alpha_1\cdots\beta_{b-1},\rho_1\cdots\sigma_{n\!-\!1}} -\mathcal{C}_{(2+b,n-1)}^{\mu\alpha\nu\beta\rho\sigma\alpha_1\cdots\beta_{b-1},\rho_1\cdots\sigma_{n\!-\!1}} \right] (k_n)_\lambda \left( \Gamma_{\rho\mu\nu} \right)^{\lambda\rho_n\sigma_n} (p_{2b-1})_\sigma (p_{2b})_\alpha (p_{2b})_\beta \\
      & \hspace{16pt} + \left[ \mathcal{C}_{(3+b,n-2)}^{\mu\nu\alpha\beta\rho\sigma\lambda\tau\alpha_1\cdots\beta_{b-1},\rho_1\cdots\sigma_{n\!-\!2}} -\mathcal{C}_{(3+b,n-2)}^{\mu\alpha\nu\beta\rho\sigma\lambda\tau\alpha_1\cdots\beta_{b-1},\rho_1\cdots\sigma_{n\!-\!2}} \right] \left( \Gamma_{\rho\mu\nu} \right)^{\lambda_1\rho_{n\!-\!1}\sigma_{n\!-\!1}}\left( \Gamma_{\lambda\alpha\beta} \right)^{\lambda_2\rho_n\sigma_n} \\
      & \hspace{24pt} \times (k_{n\!-\!1})_{\lambda_1} (k_n)_{\lambda_2} (p_{2b-1})_\sigma (p_{2b})_\tau \Bigg]  +\text{permutations} .
  \end{split}
\end{align}
In this expression, the permutation term accounts for all possible permutations of both scalars and gravitons. As noted above, we included all the terms, even though the second and fifth terms begin to contribute at $\kappa^2$ order.

\subsection{Horndeski $G_5 (\phi,X)$ class}

The last interaction class of the Horndeski gravity describes the most sophisticated non-minimal coupling involving three graviton momenta. We shall simplify it in complete analogy with the previous cases.

Firstly, we assume that the coupling function $G_5$ is smooth enough to admit the power series expansion:
\begin{align}
  G_5 = \sum\limits_{a,b} \Psi_{(a,b)} \phi^a X^b.
\end{align}
Here, $\Psi_{(a,b)}$ is a dimensional coupling with the mass dimension $4(1-b)-a-5$. Without the loss of generality, we use a single term for this expansion and consider the following action:
\begin{align}
  \begin{split}
    & \int d^4 x \sqrt{-g} \left[ G_5 G^{\mu\nu} \nabla_\mu\nabla_\nu\phi - \cfrac16\, G_{5,X} \left\{\left(\square\phi\right)^3 - 3 \square\phi \, \left(\nabla_\mu\nabla_\nu\phi\right)^2 +2 \left( \nabla_\mu\nabla_\nu \phi \right)^3 \right\} \right] \\
    & \to \int d^4 x \sqrt{-g} \, \Psi_{a,b} \, \phi^a X^{b-1} \left[ X \,G^{\mu\nu} \nabla_\mu\nabla_\nu\phi - \cfrac{b}{6} \left\{\left(\square\phi\right)^3 - 3 \square\phi \, \left(\nabla_\mu\nabla_\nu\phi\right)^2 +2 \left( \nabla_\mu\nabla_\nu \phi \right)^3 \right\} \right].
  \end{split}
\end{align}

Secondly, the action consists of two parts. The first part involves the Einstein tensor, and it takes the following form:
\begin{align}
  \begin{split}
    & \int d^4 x \sqrt{-g} \, G_{\mu\nu} \, \nabla^\mu \nabla^\nu\phi = \cfrac12\, \int d^4 x \, \sqrt{-g} \left[ g^{\mu\alpha} g^{\nu\beta} + g^{\mu\beta} g^{\nu\alpha} - g^{\mu\nu} g^{\alpha\beta} \right] R_{\mu\nu} \nabla_\alpha \nabla_\beta \phi \\
    & = \cfrac12\, \int d^4 x \sqrt{-g} \left[ g^{\mu\alpha} g^{\nu\beta} + g^{\mu\beta} g^{\nu\alpha} - g^{\mu\nu} g^{\alpha\beta} \right] g^{\rho\sigma} \left( \pd_\rho \Gamma_{\sigma\mu\nu} - \pd_\mu \Gamma_{\rho\sigma\nu} \right) \,\nabla_\alpha \nabla_\beta \phi\\
    & \hspace{8pt} + \cfrac12\, \int d^4 x \sqrt{-g} \left[ g^{\mu\alpha} g^{\nu\beta} + g^{\mu\beta} g^{\nu\alpha} - g^{\mu\nu} g^{\alpha\beta} \right] g^{\rho\sigma} g^{\lambda\tau} \left( \Gamma_{\rho\lambda\mu} \Gamma_{\sigma\tau\nu} - \Gamma_{\rho\lambda\tau} \Gamma_{\sigma\mu\nu} \right) \,\nabla_\alpha \nabla_\beta \phi \\
    & = \cfrac12\, \int d^4 x \sqrt{-g} \left[ g^{\mu\alpha} g^{\nu\beta} + g^{\mu\beta} g^{\nu\alpha} - g^{\mu\nu} g^{\alpha\beta} \right] g^{\rho\sigma} \left( \pd_\rho \Gamma_{\sigma\mu\nu} - \pd_\mu \Gamma_{\rho\sigma\nu} \right) \,\pd_\alpha \pd_\beta \phi\\
    & \hspace{8pt} - \cfrac12\, \int d^4 x \sqrt{-g} \left[ g^{\mu\alpha} g^{\nu\beta} + g^{\mu\beta} g^{\nu\alpha} - g^{\mu\nu} g^{\alpha\beta} \right] g^{\rho\sigma} g^{\lambda\tau} \left( \pd_\rho \Gamma_{\sigma\mu\nu} - \pd_\mu \Gamma_{\rho\sigma\nu} \right) \Gamma_{\lambda\alpha\beta} \pd_\tau \phi\\
    & \hspace{8pt} + \cfrac12\, \int d^4 x \sqrt{-g} \left[ g^{\mu\alpha} g^{\nu\beta} + g^{\mu\beta} g^{\nu\alpha} - g^{\mu\nu} g^{\alpha\beta} \right] g^{\rho\sigma} g^{\lambda\tau} \left( \Gamma_{\rho\lambda\mu} \Gamma_{\sigma\tau\nu} - \Gamma_{\rho\lambda\tau} \Gamma_{\sigma\mu\nu} \right) \,\pd_\alpha \pd_\beta \phi \\
    & \hspace{8pt} - \cfrac12\, \int d^4 x \sqrt{-g} \left[ g^{\mu\alpha} g^{\nu\beta} + g^{\mu\beta} g^{\nu\alpha} - g^{\mu\nu} g^{\alpha\beta} \right] g^{\rho\sigma} g^{\lambda\tau} g^{\omega\epsilon}  \left( \Gamma_{\rho\lambda\mu} \Gamma_{\sigma\tau\nu} - \Gamma_{\rho\lambda\tau} \Gamma_{\sigma\mu\nu} \right) \Gamma_{\omega\alpha\beta} \,\pd_\epsilon \phi .
  \end{split}
\end{align}
The second part also contains Christoffel symbols due to the presence of covariant derivatives and takes the following form:
\begin{align}
  \begin{split}
    & \int d^4 x \sqrt{-g} \, \left[ (\square\phi)^3 - 3 \square\phi \left( \nabla_\mu \nabla_\nu \phi \right)^2 + 2 \left( \nabla_\mu \nabla_\nu \phi \right)^3 \right] \\
    & =\int d^4 x \sqrt{-g} \, \left[ g^{\mu\nu} g^{\alpha\beta} g^{\rho\sigma} -3 \, g^{\mu\nu} g^{\alpha\rho} g^{\beta\sigma} + 2 g^{\nu\alpha} g^{\beta\rho} g^{\sigma\mu} \right] \nabla_\mu \nabla_\nu\phi \, \nabla_\alpha \nabla_\beta\phi\, \nabla_\rho \nabla_\sigma \phi \\
    & =\int d^4 x \sqrt{-g} \, \left[ g^{\mu\nu} g^{\alpha\beta} g^{\rho\sigma} -3 \, g^{\mu\nu} g^{\alpha\rho} g^{\beta\sigma} + 2 g^{\nu\alpha} g^{\beta\rho} g^{\sigma\mu} \right] \pd_\mu\pd_\nu\phi \, \pd_\alpha\pd_\beta\phi\, \pd_\rho\pd_\sigma \phi \\
    & \hspace{8pt} - 3 \int d^4 x \sqrt{-g} \left[ g^{\mu\nu} g^{\alpha\beta} g^{\rho\sigma} -3 \, g^{\mu\nu} g^{\alpha\rho} g^{\beta\sigma} + 2 g^{\nu\alpha} g^{\beta\rho} g^{\sigma\mu} \right] g^{\lambda\tau} \Gamma_{\lambda\mu\nu} \pd_\tau\phi \, \pd_\alpha\pd_\beta\phi\, \pd_\rho\pd_\sigma \phi \\
    & \hspace{8pt} + 3 \int d^4 x \sqrt{-g} \left[ g^{\mu\nu} g^{\alpha\beta} g^{\rho\sigma} -3 \, g^{\mu\nu} g^{\alpha\rho} g^{\beta\sigma} + 2 g^{\nu\alpha} g^{\beta\rho} g^{\sigma\mu} \right] g^{\lambda_1\tau_1} g^{\lambda_2\tau_2} \Gamma_{\lambda_1\mu\nu} \Gamma_{\lambda_2\alpha\beta} \pd_{\tau_1} \phi \, \pd_{\tau_2}\phi\, \pd_\rho\pd_\sigma \phi \\
    & \hspace{8pt} - \!\! \int d^4 x \sqrt{-g} \left[ g^{\mu\nu} g^{\alpha\beta} g^{\rho\sigma} -3 \, g^{\mu\nu} g^{\alpha\rho} g^{\beta\sigma} + 2 g^{\nu\alpha} g^{\beta\rho} g^{\sigma\mu} \right] g^{\lambda_1\tau_1} g^{\lambda_2\tau_2} g^{\lambda_3\tau_3} \Gamma_{\lambda_1\mu\nu} \Gamma_{\lambda_2\alpha\beta}\Gamma_{\lambda_3\rho\sigma} \pd_{\tau_1}\phi \, \pd_{\tau_2}\phi\, \pd_{\tau_3} \phi .
  \end{split}
\end{align}

These expressions result in the following sophisticated expression describing the perturbative structure of the interaction:
\begin{align}
  & \int d^4 x \sqrt{-g} \Psi_{a,b} \, \phi^a X^{b-1} \left[ X \,G^{\mu\nu} \nabla^\mu\nabla^\nu\phi - \cfrac{b}{6} \left\{\left(\square\phi\right)^2 - 3 \square\phi \, \left(\nabla_\mu\nabla_\nu\phi\right)^2 +2 \left( \nabla_\mu\nabla_\nu \phi \right)^3 \right\} \right] \\
  & = \sum\limits_{n=1}^\infty \int \prod\limits_{i=1}^n \cfrac{d^4 k_i}{(2\pi)^4} \, h_{\rho_i\sigma_i}(k_i) \prod\limits_{j=1}^{a+2b+1} \cfrac{d^4 p_j}{(2\pi)^4} \,\phi(p_j) \,(2\pi)^4 \delta\left( \sum k_i + \sum p_j \right) \nonumber \\
  & \hspace{8pt} \times \cfrac12 \, \kappa^n \Psi_{(a,b)} (-1)^{b+2} \left[ \mathcal{C}_{3+b,n-1}^{\mu\alpha\nu\beta\rho\sigma\alpha_1\cdots\beta_b,\rho_1\cdots\sigma_{n\!-\!1}} + \cdots\right] (k_n)_\lambda \left( (k_n)_\rho \left(\Gamma_{\sigma\mu\nu} \right)^{\lambda\rho_n\sigma_n} - (k_n)_\mu \left(\Gamma_{\rho\sigma\nu} \right)^{\lambda\rho_n\sigma_n} \right) \nonumber \\
  & \hspace{16pt} \times (p_1)_{\alpha_1} \cdots (p_{2b})_{\beta_b} (p_{2b+1})_\alpha (p_{2b+1})_\beta \nonumber \\
  & - \sum\limits_{n=2}^\infty \int \prod\limits_{i=1}^n \cfrac{d^4 k_i}{(2\pi)^4} \, h_{\rho_i\sigma_i}(k_i) \prod\limits_{j=1}^{a+2b+1} \cfrac{d^4 p_j}{(2\pi)^4} \,\phi(p_j) \,(2\pi)^4 \delta\left( \sum k_i + \sum p_j \right) \nonumber \\
  & \hspace{8pt} \times \cfrac12 \, \kappa^n \Psi_{(a,b)} (-1)^{b+2} \left[ \mathcal{C}_{4+b,n-2}^{\mu\alpha\nu\beta\rho\sigma\lambda\tau\alpha_1\cdots\beta_b,\rho_1\cdots\sigma_{n\!-\!2}} + \cdots\right] \nonumber \\
  & \hspace{16pt} \times \!\! (k_{n\!-\!1})_{\lambda_1} \!\!\! \left( \! (k_{n\!-\!1})_\rho \! \left(\Gamma_{\sigma\mu\nu} \right)^{\lambda_1\rho_{n\!-\!1}\sigma_{n\!-\!1}} \!-\! (k_{n\!-\!1})_\mu \! \left(\Gamma_{\rho\sigma\nu} \right)^{\lambda_1\rho_{n\!-\!1}\sigma_{n\!-\!1}} \right) \!\! (k_n)_{\lambda_2} \!\! \left(\Gamma_{\lambda\alpha\beta} \right)^{\lambda_2\rho_n\sigma_n} (p_1)_{\alpha_1} \!\! \cdots (p_{2b})_{\beta_b} (p_{2b+1})_\tau \nonumber \\
  & + \sum\limits_{n=2}^\infty \int \prod\limits_{i=1}^n \cfrac{d^4 k_i}{(2\pi)^4} \, h_{\rho_i\sigma_i}(k_i) \prod\limits_{j=1}^{a+2b+1} \cfrac{d^4 p_j}{(2\pi)^4} \,\phi(p_j) \,(2\pi)^4 \delta\left( \sum k_i + \sum p_j \right) \nonumber \\
  & \hspace{8pt} \times \cfrac12 \, \kappa^n \Psi_{(a,b)} (-1)^{b+2} \left[ \mathcal{C}_{4+b,n-2}^{\mu\alpha\nu\beta\rho\sigma\lambda\tau\alpha_1\cdots\beta_b,\rho_1\cdots\sigma_{n\!-\!2}} + \cdots\right] (k_{n-1})_{\lambda_1} (k_n)_{\lambda_2} \nonumber \\
  & \hspace{16pt} \times\left[ \left(\Gamma_{\rho\lambda\mu}\right)^{\lambda_1\rho_{n\!-\!1}\sigma_{n\!-\!1}} \left(\Gamma_{\sigma\tau\nu}\right)^{\lambda_2\rho_n\sigma_n} -\left(\Gamma_{\rho\lambda\tau}\right)^{\lambda_1\rho_{n\!-\!1}\sigma_{n\!-\!1}} \left( \Gamma_{\sigma\mu\nu} \right)^{\lambda_2\rho_n\sigma_n} \right] (p_1)_{\alpha_1}\cdots(p_{2b})_{\beta_b} (p_{2b+1})_\alpha (p_{2b+1})_\beta \nonumber \\
  & - \sum\limits_{n=3}^\infty \int \prod\limits_{i=1}^n \cfrac{d^4 k_i}{(2\pi)^4} \, h_{\rho_i\sigma_i}(k_i) \prod\limits_{j=1}^{a+2b+1} \cfrac{d^4 p_j}{(2\pi)^4} \,\phi(p_j) \,(2\pi)^4 \delta\left( \sum k_i + \sum p_j \right) \nonumber \\
  & \hspace{8pt} \times \cfrac12 \, \kappa^n \Psi_{(a,b)} (-1)^{b+2} \left[ \mathcal{C}_{5+b,n-2}^{\mu\alpha\nu\beta\rho\sigma\lambda\tau\omega\epsilon\alpha_1\cdots\beta_b,\rho_1\cdots\sigma_{n\!-\!2}} + \cdots\right] (k_{n\!-\!2})_{\lambda_3} (k_{n-1})_{\lambda_1} (k_n)_{\lambda_2}  \left(\Gamma_{\omega\alpha\beta}\right)^{\lambda_3\rho_{n\!-\!2}\sigma_{n\!-\!2}} \nonumber \\
  & \hspace{16pt} \times\left[ \left(\Gamma_{\rho\lambda\mu}\right)^{\lambda_1\rho_{n\!-\!1}\sigma_{n\!-\!1}} \left(\Gamma_{\sigma\tau\nu}\right)^{\lambda_2\rho_n\sigma_n} -\left(\Gamma_{\rho\lambda\tau}\right)^{\lambda_1\rho_{n\!-\!1}\sigma_{n\!-\!1}} \left( \Gamma_{\sigma\mu\nu} \right)^{\lambda_2\rho_n\sigma_n} \right] (p_1)_{\alpha_1}\cdots(p_{2b})_{\beta_b} (p_{2b+1})_\epsilon \nonumber \\
  & - \sum\limits_{n=0}^\infty \int \prod\limits_{i=1}^n \cfrac{d^4 k_i}{(2\pi)^4} \, h_{\rho_i\sigma_i}(k_i) \prod\limits_{j=1}^{a+2b+1} \cfrac{d^4 p_j}{(2\pi)^4} \,\phi(p_j) \,(2\pi)^4 \delta\left( \sum k_i + \sum p_j \right) \nonumber \\
  & \hspace{8pt }\times \cfrac{b}{6} \, \kappa^n \Psi_{(a,b)} (-1)^{b+2} \left[ \mathcal{C}_{2+b,n}^{\mu\nu\alpha\beta\rho\sigma\alpha_1\cdots\beta_{b-1},\rho_1\cdots\sigma_n} + \cdots\right] (p_1)_{\alpha_1}\cdots (p_{2b-2})_{\beta_{b-1}} \nonumber \\
  & \hspace{16pt} \times (p_{2b-1})_\mu (p_{2b-1})_\nu (p_{2b})_\alpha (p_{2b})_\beta (p_{2b+3})_\rho (p_{2b+3})_\sigma \nonumber \\
  & + \sum\limits_{n=1}^\infty \int \prod\limits_{i=1}^n \cfrac{d^4 k_i}{(2\pi)^4} \, h_{\rho_i\sigma_i}(k_i) \prod\limits_{j=1}^{a+2b+1} \cfrac{d^4 p_j}{(2\pi)^4} \,\phi(p_j) \,(2\pi)^4 \delta\left( \sum k_i + \sum p_j \right) \nonumber \\
  & \hspace{8pt }\times \cfrac{b}{2} \, \kappa^n \Psi_{(a,b)} (-1)^{b+2} \left[ \mathcal{C}_{3+b,n-1}^{\mu\nu\alpha\beta\rho\sigma\omega\tau\alpha_1\cdots\beta_{b-1},\rho_1\cdots\sigma_{n\!-\!1}} + \cdots\right] (k_n)_\lambda \left(\Gamma_{\omega\mu\nu}\right)^{\lambda\rho_n\sigma_n} (p_1)_{\alpha_1}\cdots (p_{2b-2})_{\beta_{b-1}} \nonumber \\
  & \hspace{16pt} \times (p_{2b-1})_\tau  (p_{2b})_\alpha (p_{2b})_\beta (p_{2b+1})_\rho (p_{2b+1})_\sigma \nonumber \\
  & - \sum\limits_{n=2}^\infty \int \prod\limits_{i=1}^n \cfrac{d^4 k_i}{(2\pi)^4} \, h_{\rho_i\sigma_i}(k_i) \prod\limits_{j=1}^{a+2b+1} \cfrac{d^4 p_j}{(2\pi)^4} \,\phi(p_j) \,(2\pi)^4 \delta\left( \sum k_i + \sum p_j \right) \nonumber \\
  & \hspace{8pt }\times \! \cfrac{b}{2} \, \kappa^n \Psi_{(a,b)} (-1)^{b+2} \! \left[ \mathcal{C}_{4+b,n-2}^{\mu\nu\alpha\beta\rho\sigma\omega_1\tau_1\omega_2\tau_2\alpha_1\cdots\beta_{b-1},\rho_1\cdots\sigma_{n\!-\!2}} \!+\! \cdots\right] \! (k_{n\!-\!1})_{\lambda_1} \! \left(\Gamma_{\omega_1\mu\nu}\right)^{\lambda_1\rho_{n\!-\!1}\sigma_{n\!-\!1}} (k_n)_{\lambda_2} \left(\Gamma_{\omega_2\alpha\beta}\right)^{\lambda_2\rho_n\sigma_n} \nonumber \\
  & \hspace{16pt} \times (p_1)_{\alpha_1}\cdots (p_{2b-2})_{\beta_{b-1}} (p_{2b-1})_{\tau_1}  (p_{2b})_{\tau_2} (p_{2b+1})_\rho (p_{2b+1})_\sigma \nonumber \\
  & + \sum\limits_{n=3}^\infty \int \prod\limits_{i=1}^n \cfrac{d^4 k_i}{(2\pi)^4} \, h_{\rho_i\sigma_i}(k_i) \prod\limits_{j=1}^{a+2b+1} \cfrac{d^4 p_j}{(2\pi)^4} \,\phi(p_j) \,(2\pi)^4 \delta\left( \sum k_i + \sum p_j \right) \nonumber \\
  & \hspace{8pt }\times \cfrac{b}{6} \, \kappa^n \Psi_{(a,b)} (-1)^{b+2} \left[ \mathcal{C}_{5+b,n-3}^{\mu\nu\alpha\beta\rho\sigma\omega_1\tau_1\omega_2\tau_2\omega_3\tau_3\alpha_1\cdots\beta_{b-1},\rho_1\cdots\sigma_{n\!-\!3}} + \cdots\right] (k_{n\!-\!2})_{\lambda_1} (k_{n\!-\!1})_{\lambda_2} (k_n)_{\lambda_3} \nonumber \\
  &\hspace{16pt} \times \left(\Gamma_{\omega_1\mu\nu}\right)^{\lambda_1\rho_{n\!-\!2}\sigma_{n\!-\!2}} \left(\Gamma_{\omega_2\alpha\beta}\right)^{\lambda_2\rho_{n\!-\!1}\sigma_{n\!-\!1}} \left(\Gamma_{\omega_3\rho\sigma}\right)^{\lambda_3\rho_n\sigma_n} \nonumber \\
  & \hspace{24pt} \times (p_1)_{\alpha_1}\cdots (p_{2b-2})_{\beta_{b-1}} (p_{2b-1})_{\tau_1} (p_{2b})_{\tau_2} (p_{2b+1})_{\tau_3} .\nonumber
\end{align}
This expression contains eight terms, and only term number five contributes at $\kappa^0$ level. Terms number one and six contribute at $\kappa^1$ level. Terms two, three, and seven contribute at $\kappa^2$ level, and only terms four and eight contribute at $\kappa^3$ level. 

We explicitly present the interaction rule expression for the sake of generality despite it taking up much space.
\begin{align}
  \begin{split}
    \\
    & \hspace{25pt}
    \begin{gathered}
      \begin{fmffile}{Diags/G5_vertex}
        \begin{fmfgraph*}(40,40)
          \fmfleft{L1,L2}
          \fmfright{R1,R2}
          \fmf{dbl_wiggly}{L1,V}
          \fmf{dbl_wiggly}{L2,V}
          \fmf{dashes}{V,R1}
          \fmf{dashes}{V,R2}
          \fmfdot{V}
          \fmffreeze
          \fmf{dots}{L1,L2}
          \fmf{dots}{R1,R2}
          \fmflabel{$\rho_1\sigma_1,k_1$}{L1}
          \fmflabel{$\rho_n\sigma_n,k_n$}{L2}
          \fmflabel{$p_1$}{R1}
          \fmflabel{$p_{a+2b+1}$}{R2}
          \fmflabel{$\Psi_{(a,b)}$}{V}
          \fmfv{$\Psi_{(a,b)}$,label.angle=90,label.dist=15pt}{V}
        \end{fmfgraph*}
      \end{fmffile}
    \end{gathered} = i\, \cfrac12\, \kappa^n \,\Psi_{(a,b)} (-1)^{b+2} (p_1)_{\alpha_1} (p_2)_{\beta_1} \cdots (p_{2b-1})_{\alpha_b} (p_{2b})_{\beta_b} \\ \\
    \times\Bigg[ & \left[ \mathcal{C}_{3+b,n-1}^{\mu\alpha\nu\beta\rho\sigma\alpha_1\cdots\beta_b,\rho_1\cdots\sigma_{n\!-\!1}} + \cdots\right] (k_n)_\lambda \left( (k_n)_\rho \left(\Gamma_{\sigma\mu\nu} \right)^{\lambda\rho_n\sigma_n} - (k_n)_\mu \left(\Gamma_{\rho\sigma\nu} \right)^{\lambda\rho_n\sigma_n} \right) (p_{2b+1})_\alpha (p_{2b+1})_\beta\\
      & - \left[ \mathcal{C}_{4+b,n-2}^{\mu\alpha\nu\beta\rho\sigma\lambda\tau\alpha_1\cdots\beta_b,\rho_1\cdots\sigma_{n\!-\!2}} + \cdots\right] (k_{n\!-\!1})_{\lambda_1} \left( \! (k_{n\!-\!1})_\rho  \left(\Gamma_{\sigma\mu\nu} \right)^{\lambda_1\rho_{n\!-\!1}\sigma_{n\!-\!1}} \!-\! (k_{n\!-\!1})_\mu \! \left(\Gamma_{\rho\sigma\nu} \right)^{\lambda_1\rho_{n\!-\!1}\sigma_{n\!-\!1}} \right) \\
      & \hspace{40pt} \times (k_n)_{\lambda_2} \left(\Gamma_{\lambda\alpha\beta} \right)^{\lambda_2\rho_n\sigma_n} (p_{2b+1})_\tau \\
      & +\left[ \mathcal{C}_{4+b,n-2}^{\mu\alpha\nu\beta\rho\sigma\lambda\tau\alpha_1\cdots\beta_b,\rho_1\cdots\sigma_{n\!-\!2}} + \cdots\right] (k_{n-1})_{\lambda_1} (k_n)_{\lambda_2} (p_{2b+1})_\alpha (p_{2b+1})_\beta\\
      & \hspace{40pt} \times \left[ \left(\Gamma_{\rho\lambda\mu}\right)^{\lambda_1\rho_{n\!-\!1}\sigma_{n\!-\!1}} \left(\Gamma_{\sigma\tau\nu}\right)^{\lambda_2\rho_n\sigma_n} -\left(\Gamma_{\rho\lambda\tau}\right)^{\lambda_1\rho_{n\!-\!1}\sigma_{n\!-\!1}} \left( \Gamma_{\sigma\mu\nu} \right)^{\lambda_2\rho_n\sigma_n} \right]\\
      & - \left[ \mathcal{C}_{5+b,n-2}^{\mu\alpha\nu\beta\rho\sigma\lambda\tau\omega\epsilon\alpha_1\cdots\beta_b,\rho_1\cdots\sigma_{n\!-\!2}} + \cdots\right] (k_{n - 2})_{\lambda_3} (k_{n-1})_{\lambda_1} (k_n)_{\lambda_2}  \left(\Gamma_{\omega\alpha\beta}\right)^{\lambda_3\rho_{n\!-\!2}\sigma_{n\!-\!2}}\\
      & \hspace{40pt} \times \left[ \left(\Gamma_{\rho\lambda\mu}\right)^{\lambda_1\rho_{n\!-\!1}\sigma_{n\!-\!1}} \left(\Gamma_{\sigma\tau\nu}\right)^{\lambda_2\rho_n\sigma_n} -\left(\Gamma_{\rho\lambda\tau}\right)^{\lambda_1\rho_{n\!-\!1}\sigma_{n\!-\!1}} \left( \Gamma_{\sigma\mu\nu} \right)^{\lambda_2\rho_n\sigma_n} \right] (p_{2b+1})_\epsilon \Bigg]\\
    & + i\, \cfrac12\, \kappa^n \,\Psi_{(a,b)} (-1)^{b+2} \, b\, (p_1)_{\alpha_1} (p_2)_{\beta_1} \cdots (p_{2b-3})_{\alpha_{b-1}} (p_{2b-2})_{\beta_{b-1}}  \Bigg[ \\
      & - \cfrac{1}{3} \left[ \mathcal{C}_{2+b,n}^{\mu\nu\alpha\beta\rho\sigma\alpha_1\cdots\beta_{b-1},\rho_1\cdots\sigma_n} + \cdots\right] (p_{2b-1})_\mu (p_{2b-1})_\nu (p_{2b})_\alpha (p_{2b})_\beta (p_{2b+1})_\rho (p_{2b+1})_\sigma \\
      & + \, \left[ \mathcal{C}_{3+b,n-1}^{\mu\nu\alpha\beta\rho\sigma\omega\tau\alpha_1\cdots\beta_{b-1},\rho_1\cdots\sigma_{n\!-\!1}} + \cdots\right] (k_n)_\lambda \left(\Gamma_{\omega\mu\nu}\right)^{\lambda\rho_n\sigma_n} (p_{2b-1})_\tau  (p_{2b})_\alpha (p_{2b})_\beta (p_{2b+1})_\rho (p_{2b+1})_\sigma \\
      & -  \left[ \mathcal{C}_{4+b,n-2}^{\mu\nu\alpha\beta\rho\sigma\omega_1\tau_1\omega_2\tau_2\alpha_1\cdots\beta_{b-1},\rho_1\cdots\sigma_{n\!-\!2}} + \cdots\right] (k_{n\!-\!1})_{\lambda_1} \left(\Gamma_{\omega_1\mu\nu}\right)^{\lambda_1\rho_{n\!-\!1}\sigma_{n\!-\!1}} (k_n)_{\lambda_2} \left(\Gamma_{\omega_2\alpha\beta}\right)^{\lambda_2\rho_n\sigma_n} \\
      & \hspace{40pt} \times(p_{2b-1})_{\tau_1}  (p_{2b})_{\tau_2} (p_{2b+1})_\rho (p_{2b+1})_\sigma \\
      & + \cfrac{1}{3} \left[ \mathcal{C}_{5+b,n-3}^{\mu\nu\alpha\beta\rho\sigma\omega_1\tau_1\omega_2\tau_2\omega_3\tau_3\alpha_1\cdots\beta_{b-1},\rho_1\cdots\sigma_{n\!-\!3}} + \cdots\right]   \left(\Gamma_{\omega_1\mu\nu}\right)^{\lambda_1\rho_{n\!-\!2}\sigma_{n\!-\!2}} \left(\Gamma_{\omega_2\alpha\beta}\right)^{\lambda_2\rho_{n\!-\!1}\sigma_{n\!-\!1}} \left(\Gamma_{\omega_3\rho\sigma}\right)^{\lambda_3\rho_n\sigma_n} \\
      & \hspace{40pt} \times (k_{n - 2})_{\lambda_1} (k_{n\!-\!1})_{\lambda_2} (k_n)_{\lambda_3} (p_{2b-1})_{\tau_1} (p_{2b})_{\tau_2} (p_{2b+1})_{\tau_3}  \Bigg] + \text{permutations}.
  \end{split}
\end{align}
As noted above, different terms in this expression contribute to different orders. We include all the terms in the expression for the sake of generality. The permutation terms account for all permutations of all $a + 2\, b +1$ scalar fields and $n$ gravitons.

\subsection{Gauss-Bonnet term}\label{Section_Gauss_Bonnet}

Concluding the discussion of Horndeski gravity, we shall touch upon the scalar field coupling to the Gauss-Bonnet term for two reasons. First, such a coupling belongs to Horndeski gravity, but its parametrisation is complicated and involves $\ln$ functions in both $G_4$ and $G_5$. Consequently, it is essential to obtain the corresponding interaction rule separately. Secondly, the coupling provides interesting phenomenology. In particular, it allows a theory to have black holes with scalar hair, which are of utmost theoretical interest. Therefore, the coupling to the Gauss-Bonnet term shall be discussed separately.

It is useful to make the structure of the Gauss-Bonnet term explicit. Sometimes, defining the term via the generalised Kronecker symbol is useful. For our purposes, on the contrary, it is more practical to make the structure of Riemann tensor indices contractions explicit:
\begin{align}
  \begin{split}
    \sqrt{-g} \, \mathcal{G} &= \sqrt{-g}\left[ R^2 - 4 R_{\mu\nu}^2 + R_{\mu\nu\alpha\beta}^2\right] \\
    &=\sqrt{-g}\left[ g^{\mu\alpha}g^{\nu\beta}g^{\sigma\tau}g^{\rho\lambda} - 4 g^{\mu\alpha}g^{\nu\sigma}g^{\beta\tau}g^{\rho\lambda} + g^{\mu\rho}g^{\nu\sigma}g^{\alpha\lambda}g^{\beta\tau} \right] R_{\mu\nu\alpha\beta} R_{\rho\sigma\lambda\tau} .
  \end{split}        
\end{align}

We introduce the $\mathfrak{T}$ tensor that describes the contraction symmetrically:
\begin{align}
  \mathfrak{T}^{\mu\nu\alpha\beta\rho\sigma\lambda\tau} = \operatorname{Symm}\left[ g^{\mu\alpha}g^{\nu\beta}g^{\sigma\tau}g^{\rho\lambda} - 4 g^{\mu\alpha}g^{\nu\sigma}g^{\beta\tau}g^{\rho\lambda} + g^{\mu\rho}g^{\nu\sigma}g^{\alpha\lambda}g^{\beta\tau} \right].
\end{align}
The symmetrisation is performed in such a way that the tensor has the following properties:
\begin{align}
  \begin{split}
    & \mathfrak{T}^{\mu\nu\alpha\beta\rho\sigma\lambda\tau} = - \mathfrak{T}^{\nu\mu\alpha\beta\rho\sigma\lambda\tau} = - \mathfrak{T}^{\mu\nu\beta\alpha\rho\sigma\lambda\tau} = - \mathfrak{T}^{\mu\nu\alpha\beta\sigma\rho\lambda\tau} = - \mathfrak{T}^{\mu\nu\alpha\beta\rho\sigma\tau\lambda}, \\
    & \mathfrak{T}^{\mu\nu\alpha\beta\rho\sigma\lambda\tau} = \mathfrak{T}^{\alpha\beta\mu\nu\rho\sigma\lambda\tau} = \mathfrak{T}^{\mu\nu\alpha\beta\lambda\tau\rho\sigma} = \mathfrak{T}^{\rho\sigma\lambda\tau\mu\nu\alpha\beta} .
  \end{split}
\end{align}
Introducing the $\mathfrak{T}$ tensor is useful because it allows us to use Riemann tensor symmetries and obtain a more compact expression.

The following formula gives the Riemann tensor with all low indices:
\begin{align}
  R_{\mu\nu\alpha\beta} = \pd_\mu \Gamma_{\alpha\nu\beta} - \pd_\nu \Gamma_{\alpha\mu\beta} + g^{\rho\sigma} \left[ \Gamma_{\rho\nu\alpha} \Gamma_{\sigma\mu\beta} - \Gamma_{\rho\mu\alpha} \Gamma_{\sigma\nu\beta} \right] .
\end{align}
It results in the following expression for this term:
\begin{align}
  \int d^4 x \sqrt{-g} \, \mathfrak{T}^{\mu\nu\alpha\beta\rho\sigma\lambda\tau} 4 \left[ \pd_\mu \Gamma_{\alpha\nu\beta} \pd_\rho \Gamma_{\lambda\sigma\tau} + 2 g^{\omega\epsilon} \pd_\mu \Gamma_{\alpha\nu\beta} \Gamma_{\omega\sigma\lambda} \Gamma_{\epsilon\rho\tau} + g^{\omega_1\epsilon_1} g^{\omega_2\epsilon_2} \Gamma_{\omega_1\nu\alpha} \Gamma_{\epsilon_1\mu\beta} \Gamma_{\omega_2\sigma\lambda} \Gamma_{\epsilon_2\rho\tau} \right] .
\end{align}

To describe a coupling of a scalar field to the Gauss-Bonnet term, we shall introduce a coupling function in full analogy with the other Horndeski interactions:
\begin{align}
  \int d^4 x \sqrt{-g} \, \mathcal{G} \, f(\phi).
\end{align}
In contrast with the previous cases, the Gauss-Bonnet term can only healthy couple the scalar field itself, but not to its kinetic term. Without the loss of generality, we assume that the coupling function is smooth enough to admit the power series expansion, so we replace it with a single term of this series:
\begin{align}
  \int d^4 x \sqrt{-g} \, \mathcal{G} \, f(\phi) \to \int d^4 x \sqrt{-g} \, \mathcal{G} \, \mathbf{g}\, \phi^q.
\end{align}
Here $q$ is a positive integer, and $\mathbf{g}$ is a dimensional coupling with the mass dimension $-q$.

Consequently, a coupling of the Gauss-Bonnet term to scalar fields has the following perturbative structure:
\begin{align}
  \begin{split}
    & \int d^4 x \sqrt{-g} \, \mathcal{G} \, \mathbf{g} \, \phi^q \\
    &= \sum\limits_{n=2}^\infty \int \prod\limits_{i=1}^n \cfrac{d^4 k_i}{(2\pi)^4} \, h_{\rho_i\sigma_i}(k_i) \prod\limits_{j=1}^q \cfrac{d^4 p}{(2\pi)^4} \, \phi(p_j) (2\pi)^4 \delta\left( \sum k_i + \sum p_j\right)\\
    & \hspace{8pt} \times \kappa^n \mathbf{g} \,4 \, \left(\sqrt{-g} \,  \mathfrak{T}^{\mu\nu\alpha\beta\rho\sigma} \right)^{\rho_1\cdots \sigma_{n\!-\!2}} (k_{n\!-1\!})_{\lambda_1} (k_n)_{\lambda_2} (k_{n\!-1\!})_\mu (k_n)_\rho \left( \Gamma_{\alpha\nu\beta} \right)^{\lambda_1\rho_{n\!-\!1}\sigma_{n\!-\!1}} \left( \Gamma_{\lambda\sigma\tau} \right)^{\lambda_2\rho_n\sigma_n} \\
    & \hspace{8pt} + \sum\limits_{n=3}^\infty \int \prod\limits_{i=1}^n \cfrac{d^4 k_i}{(2\pi)^4} \, h_{\rho_i\sigma_i}(k_i) \prod\limits_{j=1}^q \cfrac{d^4 p}{(2\pi)^4} \, \phi(p_j) (2\pi)^4 \delta\left( \sum k_i + \sum p_j\right)\\
    & \hspace{16pt} \times \kappa^n \mathbf{g} \,8  \, \left(\sqrt{-g} \,  \mathfrak{T}^{\mu\nu\alpha\beta\rho\sigma} g^{\omega\epsilon} \right)^{\rho_1\cdots \sigma_{n\!-\!3}} (k_{n\!-\!2})_{\lambda_1} (k_{n\!-\!1})_{\lambda_2} (k_n)_{\lambda_3} (k_{n\!-\!2})_\mu \\
    & \hspace{24pt} \times\left( \Gamma_{\alpha\nu\beta} \right)^{\lambda_1\rho_{n\!-\!2}\sigma_{n\!-\!2}} \left( \Gamma_{\omega\sigma\lambda} \right)^{\lambda_2\rho_{n\!-\!1}\sigma_{n\!-\!1}} \left( \Gamma_{\epsilon\rho\tau} \right)^{\lambda_3\rho_n\sigma_n} \\
    & \hspace{8pt}+ \sum\limits_{n=4}^\infty \int \prod\limits_{i=1}^n \cfrac{d^4 k_i}{(2\pi)^4} \, h_{\rho_i\sigma_i}(k_i) \prod\limits_{j=1}^q \cfrac{d^4 p}{(2\pi)^4} \, \phi(p_j) (2\pi)^4 \delta\left( \sum k_i + \sum p_j\right)\\
    & \hspace{16pt} \times \kappa^n \mathbf{g} \,4  \, \left(\sqrt{-g} \,  \mathfrak{T}^{\mu\nu\alpha\beta\rho\sigma} g^{\omega_1\epsilon_1} g^{\omega_2\epsilon_2} \right)^{\rho_1\cdots \sigma_{n\!-\!4}} (k_{n\!-\!3})_{\lambda_1} (k_{n\!-\!2})_{\lambda_2}(k_{n\!-\!1})_{\lambda_3}(k_n)_{\lambda_4} \\
    & \hspace{24pt} \times \left( \Gamma_{\omega_1\nu\alpha} \right)^{\lambda_1\rho_{n\!-\!3}\sigma_{n\!-\!3}}  \left(\Gamma_{\epsilon_1\mu\beta} \right)^{\lambda_2\rho_{n\!-\!2}\sigma_{n\!-\!2}} \left( \Gamma_{\omega_2\sigma\lambda} \right)^{\lambda_3\rho_{n\!-\!1}\sigma_{n\!-\!1}} \left( \Gamma_{\epsilon_2\rho\tau} \right)^{\lambda_4\rho_n \sigma_n} .
  \end{split}
\end{align}
In full accordance with the previous results, this interaction does not contribute at $\kappa^0$ and $\kappa^1$. The first term of this expression contributes at $\kappa^2$, the second at $\kappa^3$, and the last at $\kappa^4$.

This expression results in the following interaction rule:
\begin{align}
  \begin{split}
    \\
    & \hspace{35pt}
    \begin{gathered}
      \begin{fmffile}{Diags/GB_vertex}
        \begin{fmfgraph*}(30,30)
          \fmfleft{L1,L2}
          \fmfright{R1,R2}
          \fmf{dbl_wiggly}{L1,V}
          \fmf{dbl_wiggly}{L2,V}
          \fmf{dashes}{V,R1}
          \fmf{dashes}{V,R2}
          \fmfdot{V}
          \fmffreeze
          \fmf{dots}{L1,L2}
          \fmf{dots}{R1,R2}
          \fmflabel{$\rho_1\sigma_1,k_1$}{L1}
          \fmflabel{$\rho_n\sigma_n,k_n$}{L2}
          \fmflabel{$p_1$}{R1}
          \fmflabel{$p_q$}{R2}
          \fmflabel{$\mathbf{g}$}{V}
          \fmfv{$g$,label.angle=90,label.dist=15pt}{V}
        \end{fmfgraph*}
      \end{fmffile}
    \end{gathered}
    = i\, \kappa^n 4 \mathbf{g} \Bigg[ \!\! \left(\sqrt{-g} \mathfrak{T}^{\mu\nu\alpha\beta\rho\sigma} \right)^{\rho_1\cdots \sigma_{n\!-\!2}} \! (k_{n\!-1\!})_{\lambda_1} \! (k_n)_{\lambda_2} (k_{n\!-1\!})_\mu (k_n)_\rho \left( \Gamma_{\alpha\nu\beta} \right)^{\lambda_1\rho_{n\!-\!1}\sigma_{n\!-\!1}} \! \left( \Gamma_{\lambda\sigma\tau} \right)^{\lambda_2\rho_n\sigma_n} \\ \\
      & + \! 2  \left(\sqrt{-g} \,  \mathfrak{T}^{\mu\nu\alpha\beta\rho\sigma} g^{\omega\epsilon} \right)^{\rho_1\cdots \sigma_{n\!-\!3}} \! (k_{n\!-\!2})_{\lambda_1} \! (k_{n\!-\!1})_{\lambda_2} \! (k_n)_{\lambda_3} (k_{n\!-\!2})_\mu\left( \Gamma_{\alpha\nu\beta} \right)^{\lambda_1\rho_{n\!-\!2}\sigma_{n\!-\!2}} \left( \Gamma_{\omega\sigma\lambda} \right)^{\lambda_2\rho_{n\!-\!1}\sigma_{n\!-\!1}} \left( \Gamma_{\epsilon\rho\tau} \right)^{\lambda_3\rho_n\sigma_n} \\
      & + \left(\sqrt{-g} \,  \mathfrak{T}^{\mu\nu\alpha\beta\rho\sigma} g^{\omega_1\epsilon_1} g^{\omega_2\epsilon_2} \right)^{\rho_1\cdots \sigma_{n\!-\!4}} (k_{n\!-\!3})_{\lambda_1} (k_{n\!-\!2})_{\lambda_2}(k_{n\!-\!1})_{\lambda_3}(k_n)_{\lambda_4} \\
      & \hspace{20pt} \times \left( \Gamma_{\omega_1\nu\alpha} \right)^{\lambda_1\rho_{n\!-\!3}\sigma_{n\!-\!3}}  \left(\Gamma_{\epsilon_1\mu\beta} \right)^{\lambda_2\rho_{n\!-\!2}\sigma_{n\!-\!2}} \left( \Gamma_{\omega_2\sigma\lambda} \right)^{\lambda_3\rho_{n\!-\!1}\sigma_{n\!-\!1}} \left( \Gamma_{\epsilon_2\rho\tau} \right)^{\lambda_4\rho_n \sigma_n} \Bigg] + \text{permutations}.
  \end{split}
\end{align}
Similarly to the previous cases, the permutation term accounts for all permutations of all $n$ gravitons and $q$ scalars.

\section{Propagator of massive gravity}\label{Section_massive_gravity}

Massive gravity is a direction of research that aims to create a consistent theory of spin $2$ massive fundamental particles. The detailed review of massive gravity lies far beyond the scope of this paper and can be found in \cite{deRham:2014zqa,Hinterbichler:2011tt}. Here, we only highlight the most fundamental challenges one faces in an attempt to implement it for \texttt{FeynGrav}.

The kinetic term describing the propagation of a spin $2$ particle with a non-vanishing mass is fixed uniquely and known as the Fierz-Pauli Lagrangian:
\begin{align}
  \int d^D x \left[ -\cfrac12\, h^{\mu\nu} \mathcal{O}_{\mu\nu\alpha\beta} h^{\alpha\beta} \right].
\end{align}
Here $\mathcal{O}_{\mu\nu\alpha\beta}$ is an operator defiened as follows:
\begin{align}
  \mathcal{O}_{\mu\nu\alpha\beta} = \left[ \cfrac12 \left( \eta_{\mu\alpha}\eta_{\nu\beta} + \eta_{\mu\beta} \eta_{\nu\alpha} \right) - \eta_{\mu\nu} \eta_{\alpha\beta} \right] (\square + m^2) - \pd_\mu\pd_\alpha \eta_{\nu\beta} - \pd_\mu \pd_\beta \eta_{\nu\alpha} + \pd_\mu \pd_\nu \eta_{\alpha\beta} + \pd_\alpha \pd_\beta \eta_{\mu\nu} .
\end{align}
The part of the Fierz-Pauli action that does not depend on the mass is taken directly from general relativity, where the gauge symmetry defines it uniquely. The mass term is not fixed uniquely from the first principles since there are two invariants quadratic in perturbations and free from derivatives: $h_{\mu\nu}^2$ and $(\eta^{\mu\nu} h_{\mu\nu} )^2$. Since the theory does not admit a gauge symmetry, the relative coefficient between these terms remains free. However, the stability reasoning fixed it uniquely. Unless the mass term takes the Fierz-Pauli form, the action will contain a ghost degree of freedom.

The massive graviton propagator, consequently, is also unique:
\begin{align}
  \mathcal{G}_{\mu\nu\alpha\beta} (p) = \cfrac{i}{p^2 - m^2} \left[ \cfrac12\left(P_{\mu\alpha} P_{\nu\beta} + P_{\mu\beta} P_{\nu\alpha}\right) - \cfrac{1}{D-1}  P_{\mu\nu} P_{\alpha\beta} \right] .
\end{align}
Here, operators $P_{\mu\nu}$ are defined as follows:
\begin{align}
  P_{\mu\nu} (p) = \eta_{\mu\nu} - \cfrac{ p_\mu p_\nu }{m^2} .
\end{align}

Due to the absence of symmetry, stability is the only guiding principle that defines the interaction sector of the theory. Massive gravity models are usually plagued by the Boulware-Deser ghost \cite{Boulware:1972yco}, so the absence of the ghost is the only guiding principle.

To the best of our knowledge, the two most widely recognised consistent massive gravity models are the Dvali-Gabadadze-Porrati \cite{Dvali:2000hr,Dvali:2000rv,Dvali:2000xg} and de Rham-Gabadadze-Tolley \cite{deRham:2010kj}. The Dvali-Gabadadze-Porrati describes massive gravity in terms of a five-dimensional braneworld model with infinite extra dimensions. At the same time, The de Rham-Gabadadze-Tolley extensively relies on the vierbein formulation of gravity and bi-gravity. 

Both models use formalisms that cannot easily be converted to standard four-dimensional geometric descriptions. Without this, there is no direct way to implement them in \texttt{FeynGrav}. Further publications will address this challenge since it deserves detailed treatment.

The graviton propagator presents a notable exception since, as noted above, it is defined uniquely. Therefore, the new version of \texttt{FeynGrav} implements it directly to provide additional tools to study quantum gravity.

\section{Axion-like coupling to a single vector field}\label{Section_axion}

Quantum chromodynamics and $SU(N)$ Yang-Mills theories generally point towards the possible existence of axions. The physics and phenomenology of axions are vast and lie beyond the scope of this paper. Additionally, since there are many models of axions, their implementation in \texttt{FeynGrav} requires a mode of detailed analysis. Such an analysis, together with the most perspective models, will be discussed in the forthcoming publications. This section only discusses the simplest axion-like coupling between a scalar and a single vector field. The reason for this is twofold. Firstly, this coupling is simple enough to be implemented immediately in \texttt{FeynGrav}. Secondly, such a coupling is used in many models involving scalar fields, providing an interesting phenomenology.

Within the $SU(N)$ Yang-Mills model in a flat spacetime, the following term is a complete derivative:
\begin{align}
  \int d^4 x \,\cfrac12\, \theta \, \operatorname{tr} \left[ G_{\mu\nu} \, \widetilde{G}^{\mu\nu} \right].
\end{align}
Here, $\theta$ is an arbitrary constant, $G_{\mu\nu}$ is the Yang-Mills field tensor, and $\widetilde{G}^{\mu\nu}$ is the Hodge dual tensor:
\begin{align}
  \widetilde{G}^{\mu\nu} \overset{\text{def}}{=} \cfrac12\, \varepsilon^{\mu\nu\alpha\beta} G_{\alpha\beta} .
\end{align}
Here, $\varepsilon^{\mu\nu\alpha\beta}$ is the Levi-Chivita tensor. Since it is a complete derivative, it does not contribute to the classical field equations and interaction rules. However, it is of enormous importance for the vacuum structure of the theory \cite{Wilczek:1977pj,Weinberg:1977ma,tHooft:1976rip,Callan:1976je,Jackiw:1976pf}.

The theory cannot predict the value of the constant $\theta$. Experiments, in turn, put an extreme constraint on the coupling \cite{Baluni:1978rf,Crewther:1979pi,Ubaldi:2008nf}:
\begin{align}
  \lvert \theta \rvert < 10^{-10} .
\end{align}
One way to explain such a small value is to promote $\theta$ to a dynamical field $\theta(x)$, called an axion, with the non-minimal coupling to the gauge field:
\begin{align}
  \int d^4 x \,\cfrac12\, \theta(x) \, \operatorname{tr} \left[ G_{\mu\nu} \, \widetilde{G}^{\mu\nu} \right].
\end{align}

Models of axions within quantum chromodynamics led to many other models involving similar axion-like couplings. This paper only discusses the coupling between a scalar field and a single vector field. Despite its apparent simplicity, such a coupling has a broad phenomenology \cite{Sikivie:1983ip,Marsh:2015xka,Raffelt:2006cw}.

We adopt the following parametrisation of the axion-like coupling between a scalar field $\phi$ and a single vector field in curved spacetime:
\begin{align}
  \int d^4 x \sqrt{-g} \left[ \cfrac12\, F_{\mu\nu}\, \widetilde{F}^{\mu\nu} \, f(\phi) \right]
\end{align}
Here $f(\phi)$ is a infinitely differentiable function of the scalar field $\phi$; $F_{\mu\nu}$ is the standard field strength tensor:
\begin{align}
  F_{\mu\nu} = \nabla_\mu A_\nu - \nabla_\nu A_\mu = \pd_\mu A_\nu - \pd_\nu A_\mu ,
\end{align}
and $\widetilde{F}^{\mu\nu}$ is the dual tensor:
\begin{align}
  \widetilde{F}^{\mu\nu} \overset{\text{def}}{=} \varepsilon^{\mu\nu\alpha\beta} F_{\alpha\beta} .
\end{align}
In full similarity with the other models, one can replace the function $f$ with a single power-like term without the loss of generality. The interaction terms take the following simple form:
\begin{align}
  \int d^4 x \sqrt{-g}\, \cfrac14 \,\varepsilon^{\mu\nu\alpha\beta} \,F_{\mu\nu}\, F_{\alpha\beta} \, \cfrac{\lambda}{l!} \,\phi^l .
\end{align}
Here, $\lambda$ is a coupling with the mass dimension $-l$.

The perturbative structure of the interaction is obtained similarly to the other models discussed in this article:
\begin{align}
  \begin{split}
    &\int d^4 x \sqrt{-g}\, \cfrac14 \,\varepsilon^{\mu\nu\alpha\beta} \,F_{\mu\nu}\, F_{\alpha\beta} \, \cfrac{\lambda}{l!} \, \phi^l \\
    &=\sum\limits_{n=0}^\infty \int \prod\limits_{i=1}^n \cfrac{d^4 k_i}{(2\pi)^4} h_{\rho_i\sigma_i}(k_i) \cfrac{d^4 q_1}{(2\pi)^4} A_{\lambda_1}(q_1) \cfrac{d^4 q_2}{(2\pi)^4} A_{\lambda_2} (q_2) \prod\limits_{j=1}^l \cfrac{d^4 p_j}{(2\pi)^4} \phi(p_j) \, (2\pi)^4 \delta\left(\sum k_i + q_1 + q_2 + \sum p_j\right)\\
    & \hspace{8pt} \times (-1)\kappa^n \,\cfrac{\lambda}{l!} \, \left( \sqrt{-g} \right)^{\rho_1\sigma_1\cdots\rho_n\sigma_n} \varepsilon^{\tau_1\lambda_1\tau_2\lambda_2} (q_1)_{\tau_1} (q_2)_{\tau_2} .
  \end{split}
\end{align}
The corresponding interaction rule reads:
\begin{align}
  \nonumber \\
  \begin{gathered}
    \begin{fmffile}{Diags/Axion_Vector_vertex}
      \begin{fmfgraph*}(40,40)
        \fmfleft{L1,L2,L3}
        \fmftop{T}
        \fmfbottom{B}
        \fmfright{R0,R1,R2,R3,R4}
        \fmf{dbl_wiggly}{L1,V}
        \fmf{dbl_wiggly}{L3,V}
        \fmf{dashes}{R1,V}
        \fmf{dashes}{R3,V}
        \fmflabel{$\lambda$}{V}
        \fmfv{$\lambda$,label.angle=90,label.dist=15pt}{V}
        \fmffreeze
        \fmfdot{V}
        \fmflabel{$\rho_1\sigma_1$}{L1}
        \fmflabel{$\rho_n\sigma_n$}{L3}
        \fmf{dots}{L1,L3}
        \fmflabel{$p_1$}{R1}
        \fmflabel{$p_l$}{R3}
        \fmf{dots}{R1,R3}
        \fmf{photon}{V,R0}
        \fmf{photon}{V,R4}
        \fmflabel{$\lambda_1,q_1$}{R0}
        \fmflabel{$\lambda_2,q_2$}{R4}
      \end{fmfgraph*}
    \end{fmffile}
  \end{gathered} \hspace{30pt} = - i\, \kappa^n \, \lambda \, \left( \sqrt{-g} \right)^{\rho_1\sigma_1\cdots\rho_n\sigma_n} \varepsilon^{\tau_1\lambda_1\tau_2\lambda_2} (q_1)_{\tau_1} (q_2)_{\tau_2} . \\ \nonumber
\end{align}
In this expression, the ``permutation'' term is absent since the expression is already symmetrical. Namely, in this parametrisation, the expression is valid for coupling to any number of scalars. It is already symmetric with respect to permutations of all graviton and vector lines.

\section{Quadratic gravity}\label{Section_Quadratic_Gravity}

Quadratic gravity is a generalisation of general relativity that includes the terms quadratic in curvature:
\begin{align}
    \mathcal{A} = \int d^4 x \sqrt{-g} \left( - \cfrac{2}{\kappa^2} \right) \left[ R - \cfrac{1}{6\,m_0^2} \,R^2 + \cfrac{1}{m_2^2} \left( R_{\mu\nu}^2 - \cfrac13 \, R^2 \right)  \right] .
\end{align}
In this expression, $\kappa$ is the same gravitational coupling used throughout this paper, and $m_0$ and $m_2$ are two parameters of the theory with the dimension of mass.

The model was extensively studied in the literature throughout the years \cite{Stelle:1976gc,Stelle:1977ry,Julve:1978xn,Fradkin:1981iu,Elizalde:1994sn,Elizalde:1994av,Lavrov:2019nuz,Calmet:2018hfb,Calmet:2018rkj,Calmet:2018rkj}. The main aim of this paper is to discuss the Feynman rules obtained within the standard approach, so we do not discuss the theory in great details. Below, we briefly recall the essential features of the model.

First and foremost, the model is the most general extension of general relativity with the dimension-four operators in the following sense. There are only four dimension-four operators that include only metric and its derivatives: $R^2$, $R_{\mu\nu}^2$, $R_{\mu\nu\alpha\beta}^2$, and $\square R$. Operator $\square R$ is a complete derivative that can only influence the surface term. The Riemann tensor square operator $R_{\mu\nu\alpha\beta}^2$ is not a complete derivative, but can always be promoted to the Gauss-Bonnet term $R_{\mu\nu\alpha\beta}^2 - 4 R_{\mu\nu}^2 + R^2$. Consequently, only two dimension-four operators can meaningfully influence the model and are both present in quadratic gravity.

Secondly, the model contains new degrees of freedom: a massive scalar and a massive spin-$2$ ghost. Both perturbative and non-perturbative results show the existence of new degrees of freedom. The new higher-order operators contain additional derivatives, which influence the propagator. The propagator receives new poles, which are associated with the new degrees of freedom \cite{Accioly:2000nm}. At the same time, there is a nonlinear map between the original parametrisation of the model and the parametrisation, making the presence of new degrees of freedom explicit \cite{Hindawi:1995an}. The discussion on the physical role of these poles is extensive and lies far beyond the scope of this paper. Consequently, we do not discuss the physical meaning of these poles and their interpretation but pursue the main aim of this paper, which is the derivation of the interaction rules of the model.

Lastly, the model admits a gauge symmetry. In complete analogy with general relativity, the model describes massless spin-$2$ states which require gauge symmetry. Therefore, one shall use a gauge-fixing procedure to obtain the propagator of the model. Since the new degrees of freedom are massive, they do not admit a gauge symmetry and do not require a gauge fixing. Consequently, one can use the same gauge fixing method employed in general relativity. In the context of  \texttt{FeynGrav}, the Faddeev-Popov prescription's gauge fixing was discussed in \cite{Latosh:2023zsi}.

Let us proceed with the derivation of the interaction rules. To obtain the Feynman rules, one shall only describe the perturbative structure of two new operators. The following formulae hold for the Riemann tensor, Ricci tensor, and the scalar curvature:
\begin{align}
  \begin{split}
    R_{\mu\nu\alpha\beta} &= \pd_\mu \Gamma_{\alpha\nu\beta} - \pd_\nu \Gamma_{\alpha\mu\beta} + g^{\rho\sigma} \left[ \Gamma_{\rho\nu\alpha} \Gamma_{\sigma\mu\beta} - \Gamma_{\rho\mu\alpha} \Gamma_{\sigma\nu\beta} \right] ,\\
    R_{\mu\nu} &= g^{\alpha\beta} R_{\alpha\mu\beta\nu} = g^{\alpha\beta} \left[ \pd_\alpha \Gamma_{\beta\mu\nu} - \pd_\mu \Gamma_{\alpha\beta\nu} \right] + g^{\alpha\beta} g^{\rho\sigma} \left[ \Gamma_{\rho\alpha\mu} \Gamma_{\sigma\beta\nu} - \Gamma_{\rho\alpha\beta} \Gamma_{\sigma\mu\nu} \right], \\
    R &= g^{\mu\nu} R_{\mu\nu} =  g^{\mu\nu} g^{\alpha\beta} \pd_\mu \left[ \Gamma_{\nu\alpha\beta} - \Gamma_{\alpha\beta\nu} \right] + g^{\mu\nu} g^{\alpha\beta} g^{\rho\sigma} \left[ \Gamma_{\mu\alpha\rho} \Gamma_{\nu\beta\sigma} - \Gamma_{\mu\alpha\beta} \Gamma_{\nu\rho\sigma} \right] .
  \end{split}
\end{align}
Consequently, the following formulae describe squares of these quantities:
\begin{align}
  \begin{split}
    R^2 =& g^{\mu\nu} g^{\alpha\beta} g^{mn} g^{ab} \pd_\mu\left\{ \Gamma_{\nu\alpha\beta} - \Gamma_{\alpha\beta\nu} \right\} \pd_m \left\{ \Gamma_{n a b } - \Gamma_{a b n} \right\} \\
    & + 2\, g^{\mu\nu} g^{\alpha\beta} g^{mn} g^{ab} g^{rs} \pd_\mu\left\{ \Gamma_{\nu\alpha\beta} - \Gamma_{\alpha\beta\nu} \right\} \left\{ \Gamma_{mar} \Gamma_{nbs} - \Gamma_{mab} \Gamma_{nrs} \right\}\\
    & + g^{\mu\nu} g^{\alpha\beta} g^{\rho\sigma} g^{mn} g^{ab} g^{rs} \left\{ \Gamma_{\mu\alpha\rho} \Gamma_{\nu\beta\sigma} - \Gamma_{\mu\alpha\beta} \Gamma_{\nu\rho\sigma} \right\} \left\{ \Gamma_{mar} \Gamma_{nbs} - \Gamma_{mab} \Gamma_{nrs} \right\} ,
  \end{split}
\end{align}
\begin{align}
  \begin{split}
    R_{\mu\nu}^2 = g^{\mu\alpha} g^{\mu\beta} \Bigg[ & g^{\rho\sigma} g^{rs} \left\{  \pd_\rho \Gamma_{\sigma\mu\nu} - \pd_\mu \Gamma_{\rho\sigma\nu} \right\} \left\{  \pd_r \Gamma_{s\alpha\beta} - \pd_\alpha \Gamma_{r s\beta} \right\} \\
      & + 2\, g^{\rho\sigma} g^{rs} g^{lt} \left\{  \pd_\rho \Gamma_{\sigma\mu\nu} - \pd_\mu \Gamma_{\rho\sigma\nu} \right\} \left\{ \Gamma_{l r \alpha} \Gamma_{t s \beta} - \Gamma_{l r s} \Gamma_{t \alpha\beta} \right\} \\
      & + g^{\rho\sigma} g^{\lambda\tau} g^{rs} g^{lt} \left\{ \Gamma_{\lambda\rho\mu} \Gamma_{\tau\sigma\nu} - \Gamma_{\lambda\rho\sigma} \Gamma_{\tau\mu\nu} \right\} \left\{ \Gamma_{l r \alpha} \Gamma_{t s \beta} - \Gamma_{l r s} \Gamma_{t \alpha\beta} \right\} \Bigg],
  \end{split}
\end{align}
\begin{align}
  \begin{split}
    R_{\mu\nu\alpha\beta}^2 = g^{\mu m} g^{\nu n} g^{\alpha a} g^{\beta b}\Bigg[ & \left\{ \pd_\mu \Gamma_{\alpha\nu\beta} - \pd_\nu \Gamma_{\alpha\mu\beta} \right\} \left\{ \pd_m \Gamma_{a n b} - \pd_n \Gamma_{a m b} \right\} \\
      & + 2 g^{rs} \left\{ \pd_\mu \Gamma_{\alpha\nu\beta} - \pd_\nu \Gamma_{\alpha\mu\beta} \right\} \left\{ \Gamma_{r n a} \Gamma_{s m b} - \Gamma_{r m a} \Gamma_{s n b} \right\} \\
      & + g^{\rho\sigma} g^{rs} \left\{ \Gamma_{\rho\nu\alpha} \Gamma_{\sigma\mu\beta} - \Gamma_{\rho\mu\alpha} \Gamma_{\sigma\nu\beta} \right\} \left\{ \Gamma_{\rho\nu\alpha} \Gamma_{\sigma\mu\beta} - \Gamma_{\rho\mu\alpha} \Gamma_{\sigma\nu\beta} \right\} \Bigg].
  \end{split}
\end{align}
The perturbative structure of $R^2$ and $R_{\mu\nu}^2$ terms reads:
\begin{align}
  \begin{split}
    \int d^4 x \sqrt{-g} R^2 = &\int d^4 x \sqrt{-g}\, g^{\mu\nu} g^{\alpha\beta} g^{mn} g^{ab} \pd_\mu\left\{ \Gamma_{\nu\alpha\beta} - \Gamma_{\alpha\beta\nu} \right\} \pd_m \left\{ \Gamma_{n a b } - \Gamma_{a b n} \right\} \\
    & + 2\,\int d^4 x \sqrt{-g}\, g^{\mu\nu} g^{\alpha\beta} g^{mn} g^{ab} g^{rs} \pd_\mu\left\{ \Gamma_{\nu\alpha\beta} - \Gamma_{\alpha\beta\nu} \right\} \left\{ \Gamma_{mar} \Gamma_{nbs} - \Gamma_{mab} \Gamma_{nrs} \right\}\\
    & + \int d^4 x \sqrt{-g}\, g^{\mu\nu} g^{\alpha\beta} g^{\rho\sigma} g^{mn} g^{ab} g^{rs} \left\{ \Gamma_{\mu\alpha\rho} \Gamma_{\nu\beta\sigma} - \Gamma_{\mu\alpha\beta} \Gamma_{\nu\rho\sigma} \right\} \left\{ \Gamma_{mar} \Gamma_{nbs} - \Gamma_{mab} \Gamma_{nrs} \right\} 
  \end{split}
\end{align}
\begin{align*}
  = & \sum\limits_{n=2}^\infty \int \prod\limits_{i=2}^n \cfrac{d^4 k_i}{(2\pi)^4} h_{\rho_i\sigma_i}(k_i) (2\pi)^4 \delta\left( \sum k_i\right) \, \kappa^n \, \left(\sqrt{-g}\, g^{\mu\nu} g^{\alpha\beta} g^{mn} g^{ab}\right)^{\rho_3\sigma_3\cdots\rho_n\sigma_n} \\
  & \times (k_1)_\mu (k_1)_{\lambda_1} (k_2)_m (k_2)_{\lambda_2} \left[ \left(\Gamma_{\nu\alpha\beta}\right)^{\lambda_1\rho_1\sigma} - \left(\Gamma_{\alpha\beta\nu}\right)^{\lambda_1\rho_1\sigma_1} \right] \left[ \left(\Gamma_{n a b }\right)^{\lambda_2\rho_2\sigma_2} - \left(\Gamma_{a b n}\right)^{\lambda_2\rho_2\sigma_2} \right]\\
  & + 2\, \sum\limits_{n=3}^\infty \int \prod\limits_{i=3}^n \cfrac{d^4 k_i}{(2\pi)^4} h_{\rho_i\sigma_i}(k_i) (2\pi)^4 \delta\left( \sum k_i\right) \, \kappa^n \, \left(\sqrt{-g}\, g^{\mu\nu} g^{\alpha\beta} g^{mn} g^{ab} g^{r s}\right)^{\rho_4\sigma_4\cdots\rho_n\sigma_n}\\
  & \hspace{8pt}\times (k_1)_\mu \! (k_1)_{\lambda_1} \! (k_2)_{\lambda_2} \! (k_3)_{\lambda_3} \!\! \left[\! \left(\Gamma_{\nu\alpha\beta}\right)^{\lambda_1\rho_1\sigma_1} \!-\! \left(\Gamma_{\alpha\beta\nu}\right)^{\lambda_1\rho_1\sigma_1} \! \right] \!\! \left[\! \left(\Gamma_{mar}\right)^{\lambda_2\rho_2\sigma_2} \!\! \left(\Gamma_{nbs}\right)^{\lambda_3\rho_3\sigma_3} \!-\! \left(\Gamma_{mab}\right)^{\lambda_2\rho_2\sigma_2} \!\! \left(\Gamma_{nrs}\right)^{\lambda_3\rho_3\sigma_3} \! \right]\\
  & + \sum\limits_{n=4}^\infty \int \prod\limits_{i=4}^n \cfrac{d^4 k_i}{(2\pi)^4} h_{\rho_i\sigma_i}(k_i) (2\pi)^4 \delta\left( \sum k_i\right) \, \kappa^n \, \left(\sqrt{-g}\, g^{\mu\nu} g^{\alpha\beta} g^{\rho\sigma}  g^{mn} g^{ab} g^{r s} \right)^{\rho_5\sigma_5\cdots\rho_n\sigma_n}\\
  & \hspace{8pt}\times (k_1)_{\lambda_1} (k_2)_{\lambda_2} (k_3)_{\lambda_3} (k_4)_{\lambda_4} \left[ \left(\Gamma_{\mu\alpha\rho}\right)^{\lambda_1\rho_1\sigma_1} \left(\Gamma_{\nu\beta\sigma}\right)^{\lambda_2\rho_2\sigma_2} - \left(\Gamma_{\mu\alpha\beta}\right)^{\lambda_1\rho_1\sigma_1} \left(\Gamma_{\nu\rho\sigma}\right)^{\lambda_2\rho_2\sigma_2} \right]\\
  & \hspace{16pt} \times \left[ \left(\Gamma_{mar}\right)^{\lambda_3\rho_3\sigma_3} \left(\Gamma_{nbs}\right)^{\lambda_4\rho_4\sigma_4} - \left(\Gamma_{mab}\right)^{\lambda_3\rho_3\sigma_3} \left(\Gamma_{nrs}\right)^{\lambda_4\rho_4\sigma_4} \right].
\end{align*}
\begin{align}
  \begin{split}
    \int d^4 x \sqrt{-g} R_{\mu\nu}^2 =& \int d^4 x \sqrt{-g} \, g^{\mu\alpha} g^{\mu\beta} g^{\rho\sigma} g^{rs} \left\{  \pd_\rho \Gamma_{\sigma\mu\nu} - \pd_\mu \Gamma_{\rho\sigma\nu} \right\} \left\{  \pd_r \Gamma_{s\alpha\beta} - \pd_\alpha \Gamma_{r s\beta} \right\} \\
    & + 2\, \int d^4 x \sqrt{-g} \, g^{\mu\alpha} g^{\mu\beta} g^{\rho\sigma} g^{rs} g^{lt} \left\{  \pd_\rho \Gamma_{\sigma\mu\nu} - \pd_\mu \Gamma_{\rho\sigma\nu} \right\} \left\{ \Gamma_{l r \alpha} \Gamma_{t s \beta} - \Gamma_{l r s} \Gamma_{t \alpha\beta} \right\} \\
    & + \int d^4 x \sqrt{-g} \, g^{\mu\alpha} g^{\mu\beta} g^{\rho\sigma} g^{\lambda\tau} g^{rs} g^{lt} \left\{ \Gamma_{\lambda\rho\mu} \Gamma_{\tau\sigma\nu} - \Gamma_{\lambda\rho\sigma} \Gamma_{\tau\mu\nu} \right\} \left\{ \Gamma_{l r \alpha} \Gamma_{t s \beta} - \Gamma_{l r s} \Gamma_{t \alpha\beta} \right\} \Bigg]
  \end{split}
\end{align}
\begin{align*}
  = & \sum\limits_{n=2}^\infty \int \prod\limits_{i=2}^n \cfrac{d^4 k_i}{(2\pi)^4} h_{\rho_i\sigma_i}(k_i) (2\pi)^4 \delta\left( \sum k_i\right) \, \kappa^n \, \left( \sqrt{-g} \, g^{\mu\alpha} g^{\mu\beta} g^{\rho\sigma} g^{rs} \right)^{\rho_3\sigma_3\cdots\rho_n\sigma_n} \\
  & \times (k_1)_{\lambda_1} (k_2)_{\lambda_2} \left[ (k_1)_\rho \left(\Gamma_{\sigma\mu\nu}\right)^{\lambda_1\rho_1\sigma_1} - (k_1)_\mu \left(\Gamma_{\rho\sigma\nu}\right)^{\lambda_1\rho_1\sigma_1} \right] \left[  (k_2)_r \left(\Gamma_{s\alpha\beta}\right)^{\lambda_2\rho_2\sigma_2} - (k_2)_\alpha \left(\Gamma_{r s\beta}\right)^{\lambda_2\rho_2\sigma_2} \right]\\
  & + 2\, \sum\limits_{n=3}^\infty \int \prod\limits_{i=3}^n \cfrac{d^4 k_i}{(2\pi)^4} h_{\rho_i\sigma_i}(k_i) (2\pi)^4 \delta\left( \sum k_i\right) \, \kappa^n \, \left(\sqrt{-g} \, g^{\mu\alpha} g^{\mu\beta} g^{\rho\sigma} g^{rs} g^{lt}\right)^{\rho_4\sigma_4\cdots\rho_n\sigma_n}\\
  & \hspace{8pt }\times (k_1)_{\lambda_1} (k_2)_{\lambda_2} (k_3)_{\lambda_3}  \left[ (k_1)_\rho \left(\Gamma_{\sigma\mu\nu}\right)^{\lambda_1\rho_1\sigma_1} - (k_1)_\mu \left(\Gamma_{\rho\sigma\nu}\right)^{\lambda_1\rho_1\sigma_1} \right] \\
  & \hspace{16pt} \times \left[ \left(\Gamma_{l r \alpha}\right)^{\lambda_2\rho_2\sigma_2} \left(\Gamma_{t s \beta}\right)^{\lambda_3\rho_3\sigma_3} - \left(\Gamma_{l r s} \right)^{\lambda_2\rho_2\sigma_2} \left( \Gamma_{t \alpha\beta} \right)^{\lambda_3\rho_3\sigma_3} \right]\\
  & + \sum\limits_{n=4}^\infty \int \prod\limits_{i=4}^n \cfrac{d^4 k_i}{(2\pi)^4} h_{\rho_i\sigma_i}(k_i) (2\pi)^4 \delta\left( \sum k_i\right) \, \kappa^n \, \left(\sqrt{-g} \, g^{\mu\alpha} g^{\mu\beta} g^{\rho\sigma} g^{\lambda\tau} g^{rs} g^{lt} \right)^{\rho_5\sigma_5\cdots\rho_n\sigma_n}\\
  & \hspace{8pt}\times (k_1)_{\lambda_1} (k_2)_{\lambda_2} (k_3)_{\lambda_3} (k_4)_{\lambda_4} \left[ \left(\Gamma_{\lambda\rho\mu}\right)^{\lambda_1\rho_1\sigma_1} \left(\Gamma_{\tau\sigma\nu}\right)^{\lambda_2\rho_2\sigma_2} - \left(\Gamma_{\lambda\rho\sigma}\right)^{\lambda_1\rho_1\sigma_1} \left(\Gamma_{\tau\mu\nu}\right)^{\lambda_2\rho_2\sigma_2} \right]\\
  & \hspace{16pt} \times \left[ \left(\Gamma_{l r \alpha}\right)^{\lambda_3\rho_3\sigma_3} \left(\Gamma_{t s \beta}\right)^{\lambda_4\rho_4\sigma_4} - \left(\Gamma_{l r s}\right)^{\lambda_3\rho_3\sigma_3} \left(\Gamma_{t \alpha\beta}\right)^{\lambda_4\rho_4\sigma_4} \right].
\end{align*}

Further, let us return to the gauge fixing procedure. In the previous publication \cite{Latosh:2023zsi}, we discussed the Faddeev-Popov gauge fixing procedure for general relativity in detail, so we refrain from repeating the discussion here. To simplify the implementation of quadratic gravity within \texttt{FeynGrav}, we introduce the same de Donder gauge fixing term:
\begin{align}
  \mathcal{A}_\text{gf} = \int d^4 x \sqrt{-g} \, \cfrac{\epsilon}{2 \, \kappa^2} \, g_{\mu\nu} \Big[ g^{\alpha\beta} \Gamma^\mu_{\alpha\beta} \Big] \Big[ g^{\rho\sigma} \Gamma^\nu_{\rho\sigma} \Big] .
\end{align}
Here, $\epsilon$ is the gauge fixing parameter. 

That gauge fixing parameter contributes both to the propagator and vertices. At the same time, the Faddeev-Popov sector of the theory is not affected by the higher-order operators. From the physical point of view, this is because the new degrees of freedom are massive and do not admit any additional gauge symmetry.

The perturbative structure of this term was described in \cite{Latosh:2023zsi}, so we use the expression from this publication to obtain the perturbative structure of the model. Consequently, the quadratic gravity action has the following perturbative structure.
\begin{align}
  \mathcal{A} = \int d^4 x \sqrt{-g} \left( - \cfrac{2}{\kappa^2} \right) \left[ R - \cfrac13 \left( \cfrac{1}{2\,m_0^2} + \cfrac{1}{m_2^2} \right) R + \cfrac{1}{m_2^2} \, R_{\mu\nu}^2 + \cfrac{\epsilon}{2 \, \kappa^2} \, g_{\mu\nu} \Big[ g^{\alpha\beta} \Gamma^\mu_{\alpha\beta} \Big] \Big[ g^{\rho\sigma} \Gamma^\nu_{\rho\sigma} \Big] \right] 
\end{align}
\begin{align*}
  = & \sum\limits_{n=2}^\infty \int\prod\limits_{i=2}^n \cfrac{d^4 k_i}{(2\pi)^4} \,h_{\rho_i\sigma_i} (k_i) \, (2\pi)^4 \delta\left( \sum k_i \right) \kappa^{n-2} \\
  & \times \Bigg\{ 2\,\left( \sqrt{-g} g^{\mu\nu} g^{\alpha\beta} g^{\rho\sigma} \right)^{\rho_3\sigma_3\cdots\rho_n\sigma_n}  (k_1)_{\lambda_1} (k_2)_{\lambda_2} \\
  & \hspace{10pt} \times \left[ \left( \Gamma_{\alpha\mu\rho} \right)^{\lambda_1\rho_1\sigma_1} \left( \Gamma_{\sigma\nu\beta} \right)^{\lambda_2\rho_2\sigma_2} - \left( \Gamma_{\alpha\mu\nu} \right)^{\lambda_1\rho_1\sigma_1} \left( \Gamma_{\rho\beta\sigma} \right)^{\lambda_2\rho_2\sigma_2} -\cfrac{\epsilon}{4}  \left( \Gamma_{\mu\alpha\beta} \right)^{\lambda_1\rho_1\sigma_1} \left( \Gamma_{\nu\rho\sigma} \right)^{\lambda_2\rho_2\sigma_2} \right] \\
  & +  \cfrac{m_2^2 + 2 \, m_0^2}{3\, m_0^2\, m_2^2} \, \left(\sqrt{-g}\, g^{\mu\nu} g^{\alpha\beta} g^{mn} g^{ab}\right)^{\rho_3\sigma_3\cdots\rho_n\sigma_n}  (k_1)_\mu (k_1)_{\lambda_1} (k_2)_m (k_2)_{\lambda_2} \\
  & \hspace{10pt} \times \left[ \left(\Gamma_{\nu\alpha\beta}\right)^{\lambda_1\rho_1\sigma} - \left(\Gamma_{\alpha\beta\nu}\right)^{\lambda_1\rho_1\sigma_1} \right] \left[ \left(\Gamma_{n a b }\right)^{\lambda_2\rho_2\sigma_2} - \left(\Gamma_{a b n}\right)^{\lambda_2\rho_2\sigma_2} \right] \\
  & - \cfrac{2}{m_2^2} \left( \sqrt{-g} \, g^{\mu\alpha} g^{\mu\beta} g^{\rho\sigma} g^{rs} \right)^{\rho_3\sigma_3\cdots\rho_n\sigma_n} (k_1)_{\lambda_1} (k_2)_{\lambda_2}\\
  & \hspace{10pt} \times\left[ (k_1)_\rho \left(\Gamma_{\sigma\mu\nu}\right)^{\lambda_1\rho_1\sigma_1} - (k_1)_\mu \left(\Gamma_{\rho\sigma\nu}\right)^{\lambda_1\rho_1\sigma_1} \right] \left[  (k_2)_r \left(\Gamma_{s\alpha\beta}\right)^{\lambda_2\rho_2\sigma_2} - (k_2)_\alpha \left(\Gamma_{r s\beta}\right)^{\lambda_2\rho_2\sigma_2} \right] \Bigg\}
\end{align*}
\begin{align*}
  + & \sum\limits_{n=3}^\infty \int\prod\limits_{i=3}^n \cfrac{d^4 k_i}{(2\pi)^4} \,h_{\rho_i\sigma_i} (k_i) \, (2\pi)^4 \delta\left( \sum k_i \right) \kappa^{n-2} \, 2 \\
  & \times\Bigg\{ \cfrac{m_2^2 + 2 \, m_0^2}{3\, m_0^2\, m_2^2} \,  \left(\sqrt{-g}\, g^{\mu\nu} g^{\alpha\beta} g^{mn} g^{ab} g^{r s}\right)^{\rho_4\sigma_4\cdots\rho_n\sigma_n} (k_1)_\mu (k_1)_{\lambda_1} (k_2)_{\lambda_2} (k_3)_{\lambda_3}\\
  & \hspace{10pt} \times \left[ \left(\Gamma_{\nu\alpha\beta}\right)^{\lambda_1\rho_1\sigma_1} - \left(\Gamma_{\alpha\beta\nu}\right)^{\lambda_1\rho_1\sigma_1} \right] \left[ \left(\Gamma_{mar}\right)^{\lambda_2\rho_2\sigma_2} \left(\Gamma_{nbs}\right)^{\lambda_3\rho_3\sigma_3} - \left(\Gamma_{mab}\right)^{\lambda_2\rho_2\sigma_2} \left(\Gamma_{nrs}\right)^{\lambda_3\rho_3\sigma_3}  \right]\\
  & - \cfrac{2}{m_2^2} \left(\sqrt{-g} \, g^{\mu\alpha} g^{\mu\beta} g^{\rho\sigma} g^{rs} g^{lt}\right)^{\rho_4\sigma_4\cdots\rho_n\sigma_n} (k_1)_{\lambda_1} (k_2)_{\lambda_2} (k_3)_{\lambda_3}\\
  & \hspace{10pt} \times \left[ (k_1)_\rho \left(\Gamma_{\sigma\mu\nu}\right)^{\lambda_1\rho_1\sigma_1} - (k_1)_\mu \left(\Gamma_{\rho\sigma\nu}\right)^{\lambda_1\rho_1\sigma_1} \right]  \left[ \left(\Gamma_{l r \alpha}\right)^{\lambda_2\rho_2\sigma_2} \left(\Gamma_{t s \beta}\right)^{\lambda_3\rho_3\sigma_3} - \left(\Gamma_{l r s} \right)^{\lambda_2\rho_2\sigma_2} \left( \Gamma_{t \alpha\beta} \right)^{\lambda_3\rho_3\sigma_3} \right] \Bigg\}
\end{align*}
\begin{align*}
  + & \sum\limits_{n=4}^\infty \int\prod\limits_{i=4}^n \cfrac{d^4 k_i}{(2\pi)^4} \,h_{\rho_i\sigma_i} (k_i) \, (2\pi)^4 \delta\left( \sum k_i \right) \kappa^{n-2} \\
  & \times \Bigg\{ \cfrac{m_2^2 + 2 \, m_0^2}{3\, m_0^2\, m_2^2} \, \left(\sqrt{-g}\, g^{\mu\nu} g^{\alpha\beta} g^{\rho\sigma}  g^{mn} g^{ab} g^{r s} \right)^{\rho_5\sigma_5\cdots\rho_n\sigma_n} (k_1)_{\lambda_1} (k_2)_{\lambda_2} (k_3)_{\lambda_3} (k_4)_{\lambda_4} \\
  & \times \left[ \! \left(\Gamma_{\mu\alpha\rho}\right)^{\lambda_1\rho_1\sigma_1} \!\! \left(\Gamma_{\nu\beta\sigma}\right)^{\lambda_2\rho_2\sigma_2} \!-\! \left(\Gamma_{\mu\alpha\beta}\right)^{\lambda_1\rho_1\sigma_1} \!\! \left(\Gamma_{\nu\rho\sigma}\right)^{\lambda_2\rho_2\sigma_2} \! \right] \!\! \left[ \! \left(\Gamma_{mar}\right)^{\lambda_3\rho_3\sigma_3} \!\! \left(\Gamma_{nbs}\right)^{\lambda_4\rho_4\sigma_4} \!-\! \left(\Gamma_{mab}\right)^{\lambda_3\rho_3\sigma_3} \!\! \left(\Gamma_{nrs}\right)^{\lambda_4\rho_4\sigma_4} \! \right] \\
  & - \cfrac{2}{m_2^2}  \left(\sqrt{-g} \, g^{\mu\alpha} g^{\mu\beta} g^{\rho\sigma} g^{\lambda\tau} g^{rs} g^{lt} \right)^{\rho_5\sigma_5\cdots\rho_n\sigma_n} (k_1)_{\lambda_1} (k_2)_{\lambda_2} (k_3)_{\lambda_3} (k_4)_{\lambda_4} \\
  & \times\! \left[ \! \left(\Gamma_{\lambda\rho\mu}\right)^{\lambda_1\rho_1\sigma_1} \!\! \left(\Gamma_{\tau\sigma\nu}\right)^{\lambda_2\rho_2\sigma_2} \!-\! \left(\Gamma_{\lambda\rho\sigma}\right)^{\lambda_1\rho_1\sigma_1} \!\! \left(\Gamma_{\tau\mu\nu}\right)^{\lambda_2\rho_2\sigma_2} \right] \!\! \left[ \! \left(\Gamma_{l r \alpha}\right)^{\lambda_3\rho_3\sigma_3} \!\! \left(\Gamma_{t s \beta}\right)^{\lambda_4\rho_4\sigma_4} \!-\! \left(\Gamma_{l r s}\right)^{\lambda_3\rho_3\sigma_3} \!\! \left(\Gamma_{t \alpha\beta}\right)^{\lambda_4\rho_4\sigma_4} \! \right] \! \Bigg\} .
\end{align*}

We have two reasons for not presenting the expression of the quadratic gravity vertex. Firstly, one can extract the expression directly from the given formula. Secondly, the expression would occupy a significant amount of space without providing new information and would be challenging to present in an easily understandable format. Nonetheless, it is helpful to show the expression for the small metric perturbation propagator:
\begin{align}
  \begin{split}
    \mathcal{G}_{\mu\nu\alpha\beta}(k) =& i \Bigg[ \cfrac{1}{p^2} \left\{ -\cfrac12 \, P^0_{\mu\nu\alpha\beta} + \cfrac{2}{\epsilon} \, P^1_{\mu\nu\alpha\beta} + P^2_{\mu\nu\alpha\beta} + \left( \cfrac{4}{\epsilon} - \cfrac32 \right) \overline{P}^0_{\mu\nu\alpha\beta} - \cfrac12 \,\overline{\overline{P}}^0_{\mu\nu\alpha\beta} \right\}\\
      & \hspace{35pt} + \cfrac{1}{p^2 - m_0^2} \left\{ \cfrac12 \, P^0_{\mu\nu\alpha\beta} + \cfrac32 \, \overline{P}^0_{\mu\nu\alpha\beta} + \cfrac12\, \overline{\overline{P}}^0_{\mu\nu\alpha\beta} \right\} - \cfrac{1}{p^2 - m_2^2} \, P^2_{\mu\nu\alpha\beta} \Bigg].
  \end{split} 
\end{align}
Here $P^1$, $P^2$, $P^0$, $\overline{P}^0$, and $\overline{\overline{P}}^0$ are the Nieuwenhuizen operators \cite{VanNieuwenhuizen:1973fi} (the definitions from \cite{Accioly:2000nm} are used).

\section{\texttt{FeynGrav 3.0}}\label{Section_V3}

The new \texttt{FeynGrav} version 3.0 is published online open access \cite{FeynGrav}. It implements many new features while keeping the core of the package unchanged. 

\texttt{FeynGrav} is an extension of \texttt{FeynCalc} \cite{Mertig:1990an,Shtabovenko:2016sxi,Shtabovenko:2020gxv,Shtabovenko:2021hjx}, so it depends on it and requires it to function correctly. The main package \texttt{FeynGrav.wl} loads \texttt{FeynCalc} and imports libraries describing interaction vertices from the \texttt{Libs} directory. The package comes with some pre-generated libraries. Additional libraries can either be generated or downloaded \cite{https://doi.org/10.17632/9xrw2jjrbr.1}. The repository \cite{https://doi.org/10.17632/9xrw2jjrbr.1} contains files with the expression for the interaction rules in \texttt{FeynCalc} format generated with \texttt{FeynGrav}. The user shall place these files in directory \texttt{Rules}, after which the \text{FeynGrav} can import them with one of the \texttt{import} commands described below.

As noted in previous sections, the number of terms in a vertex typically grows as $2^n n!$ where $n$ is the number of gravitons involved in the interaction. Consequently, the libraries for interactions involving many gravitons occupy much space. The package provides essential libraries up to $\okappa{2}$ order while allowing the generation or downloading of the other required libraries.

The new version of the package allows the user to choose which libraries to import. When the package initialises, it imports only libraries for general relativity, the minimal gravitational coupling to scalars, Dirac fermions, and vectors. Tables \ref{Table_Imports_1} and \ref{Table_Imports_2} give the list of commands importing libraries. Each command has an option \texttt{printOutput} set to \texttt{False} by default. If the option is \texttt{True}, the command prints additional output describing the import operation.

\begin{table}[!ht]
  \centering
  \renewcommand{\arraystretch}{1.1}
  \begin{tabular}{|l|p{7cm}|}
    \hline
    \textbf{Command} & \textbf{Description} \\ \hline
    \texttt{importGravitons[n]} & Imports graviton vertices in general relativity up to $\okappa{n}$. If not specified, $n=2$. If the required libraries do not exist, imports up to the highest existing order. \\ \hline
    \texttt{importScalars[n]} & Imports graviton-scalar vertices for the minimal coupling up to $\okappa{n}$. If not specified, $n=2$. If the required libraries do not exist, imports up to the highest existing order. \\ \hline
    \texttt{importFermions[n]} & Imports graviton-Dirac fermion vertices up to $\okappa{n}$. If not specified, $n=2$. If the required libraries do not exist, imports up to the highest existing order. \\ \hline
    \texttt{importVectors[n]} & Imports graviton-vector vertices for massless and massive fields up to $\okappa{n}$. If not specified, $n=2$. If the required libraries do not exist, imports up to the highest existing order. \\ \hline
    \texttt{importSUNYM[n]} & Imports graviton-$SU(N)$ Yang-Mills vertices up to $\okappa{n}$. If not specified, $n=2$. If the required libraries do not exist, imports up to the highest existing order. \\ \hline
    \texttt{importAxionVectorVertex[n]} & Imports graviton-axion-vector interaction vertices up to $\okappa{n}$. If not specified, $n=2$. If the required libraries do not exist, imports up to the highest existing order. \\ \hline
  \end{tabular}
  \renewcommand{\arraystretch}{1} %
  \caption{Commands that import libraries for models with minimal coupling to gravity.}\label{Table_Imports_1}
\end{table}
\begin{table}[!ht]
  \centering
  \renewcommand{\arraystretch}{1.1}
  \begin{tabular}{|l|p{7cm}|} \hline
    \textbf{Command} & \textbf{Description} \\ \hline
    \texttt{importHorndeskiG2[]} & Imports vertices for the $G_2$ class of the Horndeski theory. \\ \hline
    \texttt{importHorndeskiG3[]} & Imports vertices for the $G_3$ class of the Horndeski theory. \\ \hline
    \texttt{importHorndeskiG4[]} & Imports vertices for the $G_4$ class of the Horndeski theory. \\ \hline
    \texttt{importHorndeskiG5[]} & Imports vertices for the $G_5$ class of the Horndeski theory. \\ \hline
    \texttt{importScalarGaussBonnet[n]} & Imports scalar-Gauss-Bonnet interaction vertices up to $\okappa{n}$. If not specified, $n=2$. If the required libraries do not exist, imports up to the highest existing order. \\ \hline
    \texttt{importQuadraticGravity[n]} & Imports graviton vertices in quadratic gravity up to $\okappa{n}$. If not specified, $n=2$. If the required libraries do not exist, imports up to the highest existing order. \\ \hline
  \end{tabular}
  \renewcommand{\arraystretch}{1} %
  \caption{Commands that import libraries for models with minimal coupling to gravity.}\label{Table_Imports_2}
\end{table}

\begin{table}[!ht]
  \centering
  \renewcommand{\arraystretch}{1.2}
  \begin{tabular}{|l|l|}
    \hline
    \textbf{Library} & \textbf{Orders} \\ \hline
    General relativity & up to $\okappa{2}$ \\ \hline
    Minimally coupled scalars & up to $\okappa{2}$ \\ \hline
    Dirac fermions & up to $\okappa{2}$ \\ \hline
    Vectors & up to $\okappa{2}$ \\ \hline
    SU(N) Yang-Mills & up to $\okappa{2}$ \\ \hline
    Horndeski G2 & for number of scalars from 3 to 6, up to $\okappa{2}$ \\ \hline
    Horndeski G3 & for number of scalars from 3 to 6, up to $\okappa{2}$ \\ \hline
    Horndeski G4 & for number of scalars from 3 to 4, up to $\okappa{2}$ \\ \hline
    Horndeski G5 & for number of scalars from 3 to 4, $\okappa{2}$ \\ \hline
    Scalar-Gauss-Bonnet & $\okappa{2}$ only \\ \hline
    Axion-Vectors & up to $\okappa{2}$ \\ \hline
    Quadratic Gravity & up to $\okappa{2}$ \\ \hline
  \end{tabular}
  \renewcommand{\arraystretch}{1}
  \caption{The list of libraries provided with \texttt{FeynGrav}.}\label{Table_Libraries}
\end{table}

\texttt{FeynGrav} includes a separate package for generating libraries, which is present in 
\begin{quote}
  \texttt{Libs/FeynGravLibrariesGenerator.wl}.
\end{quote} 
It lacks a user-friendly interface, which will be improved in forthcoming updates. The new version of \texttt{FeynGrav} enhances the performance of this package. First and foremost, it uses the relations described in Section \ref{Section_Recursive}. Secondly, the packages use the \texttt{FORM} symbolic manipulation system \cite{Vermaseren:2000nd,FORM}, designed to operate with symbolic expressions much faster than \texttt{Wolfram Mathematica}. Consequently, the package depends on \texttt{FORM}, and the library generation algorithm will not operate without it. However, if a user is not interested in generating new libraries, the package works as intended without the \texttt{FORM}. Table \ref{Table_Libraries} provides the complete list of libraries included in the package.

The package responsible for the library generation relies on the Feynman rules discussed in this and previous publications \cite{Latosh:2022ydd,Latosh:2023zsi,Latosh:2024anf}. The rules are implemented as additional supplementary packages and placed in \texttt{Rules} folder. These packages are designed to generate expressions that will later be treated with \texttt{FORM}, and we cannot recommend using them separately. Consequently, we shall not discuss them in detail except the \texttt{Nieuwenhuizen.wl}. The \texttt{Nieuwenhuizen.wl} package provides tools to operate with the Nieuwenhuizen operators and gauge projects. The package is self-contained and can be used alone with \texttt{FeynCalc}. The main \texttt{FeynGrav} package imports this package on initialisation and allows users to use Nieuwenhuizen operators.

The rules rely on supplementary packages \texttt{ITensor.wl}, \texttt{ETensor.wl},\linebreak\texttt{CETensor.wl}, and \texttt{CTensorGeneral.wl}. These packages contain a realisation of $\mathcal{C}_{(l;n)}$ tensors \eqref{the_C_l_n_tensor_definition} defined with recursive relations. The improved algorithm generates the same expressions for $\mathcal{C}_{(l;n)}$ tensors \eqref{the_C_l_n_tensor_definition}, which was verified by a direct comparison of the result of the calculations for $l=0,1,2,3$ and $n=0,1,2,3$. The $2\to 2$ on-shell tree-level graviton scattering amplitude within general relativity was calculated as an additional consistency test. This amplitude is well known in the literature and was first obtained in \cite{Sannan:1986tz}. The previous \cite{Latosh:2022ydd} and the new version of \texttt{FeynGrav} provide the same result for the amplitude matching the known expression. Based on these facts, it is safe to conclude that the new algorithms implementing the recurrent relations are working as intended.

To quantify the efficiency improvement of the new algorithms, the average calculation time of tensors $\mathcal{C}_{(l;n)}$ was studied. Each tensor was calculated $100$ times\footnote{The new version of \texttt{FeynGrav} uses the so-called memoisation. Each time it evaluates a computationally expensive function, it stores the evaluated expression in memory. When the function is called again, the program does not evaluate the function again but extracts the evaluated function from memory. Implementation of memoisation further improves the performance of \texttt{FeynGrav}. However, the memoisation was manually turned off for these particular efficiency tests, so it would not influence the testing results.} and the average computational time was obtained with the Wolfram Mathematica internal tools. The Table \ref{Table_Computation_Efficiency} shows the test results \footnote{The testing was performed on a computer running Ubuntu 24.04.1 LTS with motherboard MSI MS-7A14, processor Intel Core i7-7700K x8, and  16 GiB of RAM.}.

\begin{table}[!ht]
  \centering
  \renewcommand{\arraystretch}{1.2}
  \begin{tabular}{|l|l|l|l|l|l|}
    \hline
    $\mathcal{C}_{(l;n)}$ & Previous & Recursive & $\mathcal{C}_{(l;n)}$ & Previous & Recursive \\
    & algorithm, sec & algorithm, sec & & algorithm, sec & algorithm, sec \\ \hline
    $\mathcal{C}_{(0;1)}$ & 0.014 & 0.000092 & $\mathcal{C}_{(2;1)}$ & 0.000019 & 0.00017 \\ \hline
    $\mathcal{C}_{(0;2)}$ & 0.018 & 0.00067 & $\mathcal{C}_{(2;2)}$ & 0.093 & 0.0013 \\ \hline
    $\mathcal{C}_{(0;3)}$ & 0.059 & 0.0029 & $\mathcal{C}_{(2;3)}$ & 0.9 & 0.0067 \\ \hline
    $\mathcal{C}_{(0;4)}$ & 0.81 & 0.016 & $\mathcal{C}_{(2;4)}$ & 27.04 & 0.057 \\ \hline
    $\mathcal{C}_{(1;1)}$ & 0.013 & 0.00011 & $\mathcal{C}_{(3;1)}$ & 0.014 & 0.00022 \\ \hline
    $\mathcal{C}_{(1;2)}$ & 0.06 & 0.0010 & $\mathcal{C}_{(3;2)}$ & 0.12 & 0.0020 \\ \hline
    $\mathcal{C}_{(1;3)}$ & 0.53 & 0.0040 & $\mathcal{C}_{(3;3)}$ & 1.22 & 0.010 \\ \hline
    $\mathcal{C}_{(1;4)}$ & 21.9 & 0.031 & $\mathcal{C}_{(3;4)}$ & 33.93 & 0.089 \\ \hline
  \end{tabular}
  \renewcommand{\arraystretch}{1}
  \caption{Average computational time test for recursive algorithms.}\label{Table_Computation_Efficiency}
\end{table}

The new \texttt{FeynGrav} version excludes some commands describing propagators. Previous versions employed commands realising propagators with \texttt{FAD} and \texttt{SPD} commands. The \texttt{FAD} command is more appropriate for loop calculations since \texttt{FeynCalc} algorithms recognise it, while \texttt{SPD} is more suitable for tree-level calculations. The present version removed this feature since it is excessive. The \texttt{FeynCalc} has a command \texttt{FeynAmpDenominatorExplicit} that allows one to convert \texttt{FAD} to \texttt{SPD}, making the discussed commands obsolete. Table \ref{Table_Propagators} presents the list of all propagators implemented in \texttt{FeynGrav}.

\begin{table}[!ht]
  \centering
  \renewcommand{\arraystretch}{1.2}
  \begin{tabular}{|l|p{7cm}|}
    \hline
    \textbf{Command} & \textbf{Description} \\ \hline
    \texttt{ScalarPropagator} & Propagator of a scalar. \\ \hline
    \texttt{ProcaPropagator} & Propagator of a massive vector field. \\ \hline
    \texttt{GravitonPropagator} & Graviton propagator.\\ \hline
    \texttt{GravitonPropagatorMassive} & Massive graviton propagator. \\ \hline
    \texttt{QuadraticGravityPropagator} & Graviton propagator within quadratic gravity. \\ \hline
  \end{tabular}
  \renewcommand{\arraystretch}{1}
  \caption{The list of all commands for propagators.}\label{Table_Propagators}
\end{table}

The latest version includes a variety of commands that describe gravitational interaction vertices. Table \ref{Table_Vertices} presents the complete list of these commands. The package also offers graviton polarisation operators. However, the previous publication \cite{Latosh:2023zsi} discuss them in detail, so we will not repeat the discussion here.

\begin{table}[!ht]
  \centering
  \renewcommand{\arraystretch}{1.2}
  \begin{tabular}{|l|p{7cm}|}
    \hline
    \textbf{Command} & \textbf{Description} \\ \hline
    \texttt{GravitonScalarVertex} & Vertex for a scalar field kinetic energy. \\ \hline
    \texttt{GravitonScalarPotentialVertex} & Vertex for a scalar field potential. \\ \hline
    \texttt{GravitonFermionVertex} & Vertex for a Dirac fermion. \\ \hline
    \texttt{GravitonMassiveVectorVertex} & Vertex for a massive vector field. \\ \hline
    \texttt{GravitonVectorVertex} & Vertex for a massless vector field. \\ \hline
    \texttt{GravitonVectorGhostVertex} & Vertex for the Faddeev-Popov ghost for a massless vector field. \\ \hline
    \texttt{GravitonGluonVertex} & Vertex for $2$, $3$, and $4$ gluons. \\ \hline
    \texttt{GravitonQuarkGluonVertex} & Vertex for the quark-gluon interaction. \\ \hline
    \texttt{GravitonYMGhostVertex} & Vertex for the Faddeev-Popov ghost for gluons. \\ \hline
    \texttt{GravitonGluonGhostVertex} & Vertex for the gluon Faddeev-Popov ghost interaction with gluons. \\ \hline
    \texttt{GravitonVertex} & Vertex for graviton interaction within general relativity. \\ \hline
    \texttt{GravitonGhostVertex} & Vertex for the Faddeev-Popov ghost for general relativity. \\ \hline
    \texttt{GravitonAxionVectorVertex} & Vertex for the axion-like coupling of a scalar field to a single vector field. \\ \hline
    \texttt{HorndeskiG2} & Vertex for Horndeski $G_2$ interaction. \\ \hline
    \texttt{HorndeskiG3} & Vertex for Horndeski $G_3$ interaction. \\ \hline
    \texttt{HorndeskiG4} & Vertex for Horndeski $G_4$ interaction. \\ \hline
    \texttt{HorndeskiG5} & Vertex for Horndeski $G_5$ interaction. \\ \hline
    \texttt{ScalarGaussBonnet} & Vertex for scalar-Gauss-Bonnet interaction between scalars and gravitons. \\ \hline
    \texttt{QuadraticGravityVertex} & Vertex for graviton interaction within quadratic gravity. \\ \hline
  \end{tabular}
  \renewcommand{\arraystretch}{1}
  \caption{The list of all commands describing interaction vertices.}\label{Table_Vertices}
\end{table}

In conclusion, the latest version of \texttt{FeynGrav} extends the toolkit for perturbative quantum gravity. The previous version of \texttt{FeynGrav} was sufficient to study quantum gravitational effects within the standard model. The new version extends its functionality for a class of modified gravity models.

\section{Conclusions and discussion}\label{Conclusions}

This paper presents the new version of \texttt{FeynGrav} \cite{FeynGrav} and develops the theoretical framework behind it \cite{Latosh:2022ydd,Latosh:2023zsi,Latosh:2024anf}. This paper discusses the following issues and their role in the computational approach to perturbative quantum gravity.

First and foremost, we discussed recurrent relations between tensors describing the perturbative structure of the theory. From the technical point of view,  the perturbative structure of quantum gravity reduces to the inverse metric $g^{\mu\nu}$ and the volume factor $\sqrt{-g}$. We established recurrent relations for the perturbative structure of these quantities (see also discussion in \cite{Latosh:2024anf}). The recurrent relations for $g^{\mu\nu}$ have a pure algebraic. Due to the scaling of small metric perturbations, we derived the recurrent relations for the volume factor $\sqrt{-g}$. The discussion of the fundamental role of this scaling and its implication lies far beyond the scope of this paper. On the practical ground, the discussed recurrent relations improve computational efficiency since they involve much fewer operations. These recurrent relations are implemented in the current version of FeynGrav, allowing it to extend its applicability to new orders of perturbation theory.

Secondly, we discussed the Horndeski gravity. The theory describes the most general class of scalar-tensor models of gravity that admit second-order field equations. The second order of the field equations ensures that the theory is free from ghost degrees of freedom typically present in higher derivative theories. The presence of healthy higher-order interactions makes this class of models an essential object of study. On the practical ground, the theory reduces to a few special cases, and we obtained the Feynman rules for all of them. We also separately treat a scalar field coupling to the Gauss-Bonnet term. Such interaction belongs to the Horndeski theory, but within the standard parameterisation, it has a form not suitable for direct implementation within a computational package. Because of this, we used a different parameterisation of this interaction and obtained the corresponding interaction rule.

Thirdly, we addressed the axion-like coupling. Axion physics is a separate branch of studies addressing the problem of the $\theta$-term of quantum chromodynamics. The development of axion physics gave rise to many models involving axion-like coupling between a scalar and a vector field. We addressed the simplest yet intensively studied case of an axion-like coupling between a scalar field, a single massless vectors field, and gravity. The more general case of a coupling to the $SU(N)$ gauge vector field will be addressed in further publications. The discussed case is a necessary step in \texttt{FeynGrav} development, providing grounds for further implementation of axion-like interactions.

Further, we implemented the massive gravity propagator. Massive gravity can be viewed as one of the simplest extensions of general relativity. The propagator of massive gravity is well-known in literature and implemented in \texttt{FeynGrav} directly. Further implementation of massive gravity will be discussed elsewhere since it faces specific challenges. Massive gravity does not admit the gauge symmetry, so it is impossible to fix its Lagrangian uniquely. At the same time, describing the theory purely in geometrical terms is challenging, so the formalism used in \texttt{FeynGrav} may not directly apply to that case.

Lastly, we considered the quadratic gravity. The theory is widely studied because it provides a unique example of a renormalisable gravity model. At the same time, the model admits a massive spin-$2$ ghost degree of freedom. The theory is given by a Lagrangian that involves the scalar curvature squared and the Ricci tensor squared terms. The formalism used in \texttt{FeynGrav} allows one to obtain the interaction roles for this model since it is given entirely in geometrical terms. Although such formulae contain many terms, the derivation of Feynman's rules presented no fundamental challenge. The corresponding interaction rules are implemented in the latest version of the package.

The new version of \texttt{FeynGrav} provides another essential step in developing tools for perturbative quantum gravity. The previous version was sufficient to study quantum gravitational effects within the standard model. The new version extends \texttt{FeynGrav} applicability to modified gravity. The interaction rules for Horndeski gravity effectively allow one to study scalar-tensor gravity and all models that can be mapped of scalar-tensor gravity. The axion-like coupling provides a tool to study many gravity models with such coupling, which was challenging before. Implementing the massive graviton propagator allows one to address the simple scattering processes in massive gravity. Lastly, the implementation of quadratic gravity gives yet another tool to address the high-energy behaviour of that model. 

Despite all the discussed developments, there are still challenges to be addressed. The main challenge of perturbative quantum gravity is the complexity growth. Since the perturbative quantum gravity operates within the standard quantum field theory paradigm, the number of terms in an interaction vertex typically grows with the number of involved gravitons $n$ as $2^n \, n!$ since the expression shall enjoy certain symmetries. Such fast growth is an essential feature of the quantum field theory and cannot be removed from this approach. A growing number of terms is required for more efficient computations. The current version begins to address this challenge. We employed \texttt{FORM} to generate expressions for the interaction vertices. However, new tools capable of efficiently operating with higher-order expressions within the perturbative quantum gravity are yet to be developed.

One shall note that many contemporary theoretical tools make the computation of scattering amplitudes (on-shell matrix elements) more efficient (including but far from being limited to \cite{Travaglini:2022uwo,Cheung:2017kzx,Rafie-Zinedine:2018izq,Elvang:2013cua,Prinz:2020nru}). Some of such tools found their implementation in computational tools, for instance, \cite{Stelzer:1994ta,Kanaki:2000ey,Badger:2010nx,Bourjaily:2023uln,Hagiwara:2024xdh}. Some of these theoretical tools may be implemented in \texttt{FeynGrav} without altering its structure and the scope of its applicability. However, the main aim of \texttt{FeynGrav} is to provide a user with a tool to operate with Feynman rules, not with a tool to calculate on-shell amplitudes. This design choice naturally limits the package's computational efficiency and applicability to the low multiplicity amplitudes. Nonetheless, further ways to improve its multiplicity will be searched.

Additionally, we will continue to implement new gravity models within \texttt{FeynGrav}. The models involving the $\theta$-term from the quantum chromodynamics shall be implemented in further versions, allowing the study of the quantum gravity role in axion physics. Similarly, we intend to implement massive gravity models, although that may require extending the theoretical framework beyond its current state.

In summary, \texttt{FeynGrav v3.0} is the next step in developing computational tools for perturbative quantum gravity. It implements new essential models and extends the package's applicability. Further development will provide more tools to study quantum gravity and a more comprehensive range of models.

\section*{Acknowledgement}
This work was supported by the Institute for Basic Science Grant IBS-R018-Y1.

\bibliographystyle{unsrturl}
\bibliography{FG3.bib}

\end{document}